\documentclass[12pt,preprint]{aastex}             	
\usepackage{epsfig}

\newcommand{\km}{${\rm km\,s}^{-1}$}
\newcommand{\fuse}{{\em FUSE}}

\slugcomment{Accepted for publication in the ApJ}
\shortauthors{Lehner et al.}
\shorttitle{\ion{C}{2} Radiative Cooling of the Diffuse Gas}
\begin{document}

\title{\ion{C}{2} Radiative Cooling of the Diffuse Gas in the Milky Way}

\author{N.\ Lehner, B.\ P.\ Wakker, B.\ D.\ Savage}

\affil{Department of Astronomy, University of Wisconsin, 475 North Charter Street, Madison, WI 53706}

\begin{abstract}
The heating and cooling of the interstellar medium (ISM) allow
the gas in the ISM to coexist at very different temperatures in thermal
pressure equilibrium. The rate at which the gas cools or heats
is therefore a fundamental ingredient for any theory of the ISM. The heating cannot 
be directly determined, but the cooling can be inferred from observations of 
\ion{C}{2*}, which is an important coolant in different environments.  
The amount of cooling can be measured 
through either the intensity of the 157.7 $\micron$ [\ion{C}{2}] 
emission line or the \ion{C}{2*} absorption lines at 1037.018 \AA\
and 1335.708 \AA, observable with the {\em Far Ultraviolet Spectroscopic Explorer} 
and the Space Telescope Imaging Spectrograph  onboard of the
{\em Hubble Space Telescope}, respectively. 
We present the results of a survey of these far-UV absorption lines in  43 objects situated 
at $\mid\! b\! \mid \ga 30 \degr$. 
Measured column densities of  \ion{C}{2*}, \ion{S}{2}, \ion{P}{2}, and \ion{Fe}{2} 
are combined with \ion{H}{1} 21-cm emission measurements
to derive the cooling rates (per H atom using \ion{H}{1} and per nucleon
using \ion{S}{2}), and to analyze the
ionization structure, the depletion, and metallicity content
of the low-, intermediate-, and high-velocity clouds 
(LVCs, IVCs, and HVCs) along the different sightlines.
Based on the depletion and the ionization structure, 
the LVCs, IVCs, and HVCs consist mostly of warm neutral and ionized clouds.
 
For the LVCs, the mean cooling rate in erg\,s$^{-1}$ per H atom is
$-25.70^{+0.19}_{-0.36}$ dex ($1\sigma$ dispersion). With a smaller
sample and a bias toward high \ion{H}{1} column density, 
the cooling rate per nucleon is similar. The corresponding total Galactic 
\ion{C}{2} luminosity in the 157.7 $\micron$ emission line
is $L \sim 2.6 \times 10^7$ L$_\odot$. Combining $N($\ion{C}{2*}) with 
the intensity of H$\alpha$ emission, we derive that
$\sim$50\% of the \ion{C}{2*} radiative cooling comes from the warm ionized medium (WIM).
The large dispersion in the cooling rates is certainly due to a combination
of differences in the ionization fraction, in the dust-to-gas fraction, and physical conditions
between sightlines. For the IVC IV Arch at $z\!\sim\! 1$ kpc 
we find that on average the cooling is a factor 2 lower than in the LVCs
that probe gas at lower $z$. For an    
HVC (Complex C, at $z > 6$ kpc) we find the much lower rate of $-26.99 \,^{+0.21}_{-0.53}$ dex,
similar to the rates observed in a sample of damped Ly$\alpha$ absorber systems (DLAs).
The fact that in the Milky Way a substantial fraction of the \ion{C}{2} cooling 
comes from the WIM implies that this is probably also true in the DLAs.  

We also derive the electron density, assuming a typical 
temperature of the warm gas of 6000 K: 
For the LVCs, $\langle  n_e \rangle = 0.08 \pm 0.04 $ cm$^{-3}$ and 
for the IV Arch, $\langle  n_e \rangle = 0.03 \pm 0.01 $ cm$^{-3}$
($1\sigma$ dispersion). 

Finally, we measured the column densities $N($\ion{S}{2}) and $N($\ion{P}{2})
in many sightlines, and confirm that sulphur appears undepleted in the ISM. 
Phosphorus becomes progressively more deficient when $\log N($\ion{H}{1}$)> 19.7$ dex,
which can either mean that P becomes more depleted into dust as more neutral gas is present,
or that P is always depleted by about $-0.3$ dex, but the
higher value of \ion{P}{2} at lower \ion{H}{1} column density indicates the need
for an ionization correction.

\end{abstract}


\section{Introduction}\label{intro}

The diffuse interstellar medium (ISM) consists of several coexisting gas-phases: a cold neutral medium (CNM)
at $T\sim 100$ K, a warm neutral or ionized diffuse medium (WNM, WIM)  at $T \sim 6\times 10^3- 10^4$ K,
and a hot, ionized phase at $T \sim 10^6$ K. In a multi-phase medium, the CNM and the WNM
can coexist in thermal pressure equilibrium at very different temperatures because 
of the heating and the cooling properties of the gas \citep{field69,mo77,mckee95,wolfire95}. 
The heating of the gas can not be directly estimated,  but can be inferred 
from \ion{C}{2} radiative cooling \citep{pottasch79} because the
157.7 $\micron$ emission-line transition of \ion{C}{2}
is a major coolant of the interstellar gas in a wide range of environments  
\citep[][and references therein]{dalgarno72,heiles94,wolfire95,wolfire03}.
\ion{C}{2} 157.7 $\micron$ emission is a major coolant because: 
(i) After oxygen, carbon is the most abundant gas phase metal in the Universe; 
(ii) its singly-ionized form is the most abundant ionization stage in 
the diffuse CNM, WNM,  and WIM; 
(iii) the  $^2$P$_{3/2}$ fine-structure state of \ion{C}{2} is easily 
excited ($h\nu/k = 91 $ K) under typical conditions in the  CNM, WNM, and  WIM.
The amount of \ion{C}{2} radiative cooling can be determined either by measuring the \ion{C}{2*} 
absorption lines originating in the $^2$P$_{3/2}$ level at 1037.018 \AA\ and 1335.708 \AA\ in the far-ultraviolet (FUV) 
\citep[e.g.,][]{pottasch79,grewing81,gry92}
or by measuring  the [\ion{C}{2}]  $^2$P$_{3/2}\rightarrow ^2$P$_{1/2}$  157.7 $\micron$ 
line intensity in the far infra-red (FIR)
\citep[e.g.,][]{shibai91,bock93,bennett94}. Not only is the \ion{C}{2} radiative cooling
important for understanding the energy budget of the ISM, but it has also been used  
to evaluate the star formation rate (SFR) in nearby star-forming  
galaxies \citep{pierini01} and in damped Ly\,$\alpha$ systems
\citep{wolfe03a,wolfe03b}.

In our Galaxy, there are many FUV \ion{C}{2*} measurements 
for gas within a few hundred pc \citep{gry92,lehner03}, but there are only a 
few observations of more distant gas at high galactic-latitudes 
\citep[e.g,][]{savage93,spitzer93,spitzer95,savage96,fitzpatrick97}.
Recent observations of a large number ($> 100$) of high Galactic latitude
stars and extragalactic objects  
with the Space Telescope Imaging Spectrograph (STIS)
onboard of the {\em Hubble Space Telescope} ({\em HST}) 
and the {\em Far-Ultraviolet Spectroscopic 
Explorer} ({\em FUSE}) now make possible an extensive study
of the \ion{C}{2} radiative cooling in the gas 
of our Galaxy at high Galactic latitudes.

We present  the cooling rate inferred from 
the \ion{C}{2*} absorption lines at 1335.708 \AA\ and/or  1037.018 \AA\
(available in the STIS and {\em FUSE}\/ wavelength bands, respectively)
measured in the FUV continuum of 43  objects at $\mid\!  b\!  \mid  \ga 30 \degr$. 
The cooling rates of the low-, intermediate- and high-velocity gas 
in these sightlines are 
derived. In \S~\ref{elecdens}, we discuss the physical processes and the assumptions 
that are necessary to derive
the cooling rate using the ratios of the column densities of
\ion{C}{2*} to \ion{H}{1} and \ion{S}{2}.  \S~\ref{obs}
presents  the observations, and \S~\ref{anal} the analysis. In particular, we
discuss the use of a curve-of-growth analysis of the \ion{Fe}{2} absorption 
lines to derive reliable column densities for the other ions. 
In \S~\ref{multicomp}, we discuss the physical properties of the clouds
encountered in this work. 
The cooling rates are presented and analyzed in \S~\ref{coolmeas}, where we also
compare our results to more local FUV measurements and FIR studies of the 
Milky Way. In \S~\ref{otherimplication}, we present some direct implications
of the \ion{C}{2*} absorption, including the origin of \ion{C}{2*} 
absorption at high Galactic latitude (\S~\ref{lvcdiscuss}), the total \ion{C}{2} 
integrated cooling rate of the Galaxy in the warm neutral
and warm ionized gas (\S~\ref{intcool}), the electron density (\S~\ref{elmeas}),
and the comparison of our survey to high redshift studies (\S~\ref{dlacomp}).
Finally we summarize our results in \S~\ref{summary}.

\section{Electron Density and \ion{C}{2} Cooling Rate from Measurements 
of $N($\ion{C}{2*})}\label{elecdens}
The ground state of \ion{C}{2} is split into two 
fine structure levels, designated (2p) $^2$P$_{3/2}$ and (2p)
$^2$P$_{1/2}$. Transitions from the $^2$P$_{1/2}$ ground state to the $^2$D and 
$^2$S levels produce absorption lines at 1334.532 \AA\ and 1036.337 \AA,
respectively, while absorption from the excited $^2$P$_{3/2}$
state produces lines at 1335.708 \AA\ (with the oscillator
strength $f = 1.28\times 10^{-1}$) and  1037.018 \AA\ ($f = 1.18 \times  10^{-1}$). These  
transitions are directly observable with STIS and {\fuse}, 
respectively. In this work, the adopted atomic parameters
and in particular the oscillator strengths are from the updated atomic 
compilation of \citet{morton03}, except as otherwise stated.

The population of the fine-structure levels of \ion{C}{2} is governed
by the equation of excitation equilibrium: 
\begin{equation}\label{elfulleqt}
n({\rm C}^{+*}) [A_{21} + n_e \gamma_{21}(e)+ n({\rm H}) \gamma_{21}({\rm H})] = 
n({\rm C}^{+}) [n_e \gamma_{12}(e) + n({\rm H}) \gamma_{12}({\rm H}) ]
\end{equation}
where $\gamma$ are the excitation (``12") and deexcitation 
(``21") rates due to collisions with electrons and neutral 
hydrogen atoms \citep{spitzer78}, the radiative-decay probability
for the upper level is $A_{21} = 2.29\times 10^{-6}$ s$^{-1}$ \citep{nussbaumer81}, and 
$n$ is the number density in cm$^{-3}$ of the electron ($n_e$) or 
a given species $X$ ($n(X)$) with $n({\rm C}^{+*})$ referring to the 
$^2$P$_{3/2}$ state and $n({\rm C^+})$ to the $^2$P$_{1/2}$ state.
The collisional excitation rate for electron  collisions
is given by \citep{spitzer78,osterbrock89},
\begin{equation}\label{gamma}
\gamma_{12}(e) = \frac{8.63 \times 10^{-6}}{g_1 \sqrt{T}} \Omega_{12} e^\frac{-E_{12}}{kT} \, {\rm cm^3\,s^{-1}}\,;
\end{equation}
where, for \ion{C}{2}, $g_1 = 2$, $E_{12}/k = 91.7$ K, and the collision strength $\Omega_{12} $ 
varies from 1.87 to 2.90 when the temperature varies from $10^3$ to $10^4$ K \citep{hayes84}. 
The collisional excitation and deexcitation rates are related through 
$\gamma_{12} = (g_2/g_1)  \exp(-h\nu_{12}/kT) \gamma_{21} $ (where $g_2=4$).
The collisional deexcitation rate for H atoms is $\gamma_{12}({\rm H}) \sim 10^{-9}$ cm$^3$\,s$^{-1}$, roughly constant
with T \citep{york79,hollenbach89}. 
We neglect the collisional excitation 
with protons because it is inhibited by the Coulomb repulsion when 
$T < 4 \times 10^4$ K \citep{bahcall68}. We will argue in \S~\ref{multicomp}
that our sightlines are not composed of CNM but of a mixture of WNM and 
WIM, so that collisions between 
H$_2$ and \ion{C}{2} are also negligible. From the ratio 
$n_e \gamma_{12}(e)/(n({\rm H})\gamma_{12}({\rm H}) ) \approx 100 n_e/n({\rm H})  T_4^{-0.5} $ 
($T_4 = T/10^4$), one can readily see that collisional 
excitations with neutral hydrogen atoms can be neglected
for $n_e/n({\rm H}) >  0.01 $. That is for a typical  $n_e \approx 0.08$ cm$^{-3}$ (see \S~\ref{elmeas}),
$n({\rm H}) \la 10$ cm$^{-3}$. At higher $n({\rm H})$ 
our sightlines would be dominated by clouds with $T < 100$ K, but 
we discuss in \S~\ref{multicomp} that this can not be the case,
and thus we can neglect collisional excitations by hydrogen for our sightlines.
\ion{C}{2} radiative
cooling appears to be also more effective in the WIM than the pure WNM: \citet{wolfire95}
predicted a factor $\sim$10 times higher cooling rate per hydrogen nucleus 
in the WIM than in the WNM.

Thus, in the warm neutral and ionized gas, the upper level of \ion{C}{2} is excited 
by collisions with electrons  and
followed by the spontaneous emission of a 157.7 $\micron$ photon when the 
ion decays to the ground state. Eq.~\ref{elfulleqt} can then be simplified to:
\begin{equation}\label{elredeqt}
n_e \simeq 0.531 \frac{\sqrt{T}}{\Omega_{12}} e^\frac{91.7 {\rm K}}{T} \, 
		\frac{N({\rm C}^{+*})}{N({\rm C}^{+})} \, {\rm cm}^{-3}\,,
\end{equation}
where we have approximated 
$n($\ion{C}{2*}$)/n($\ion{C}{2}) by $N($\ion{C}{2*}$)/N($\ion{C}{2}).
At $T=6000 $ K, $\Omega_{12} = 2.73$ \citep{hayes84} and
 $n_e \simeq 15.29 N({\rm C}^{+*})/N({\rm C}^{+})$ cm$^{-3}$. 
  
The energy loss from spontaneous emission at 157.7 $\micron$
is in the warm-ionized and neutral-diffuse gas, 
in which the electron and hydrogen deexcitations are negligible
compared to the spontaneous radiative deexcitation ($n_e \la 1 $ cm$^{-3}$ and $n_e/n({\rm H}) >  0.01 $):
$$
\Lambda({\rm C}^+) = h\nu_{21} A_{21} n({\rm C}^{+*}) = 2.89 \times 10^{-20} n({\rm C}^{+*}) \,\,{\rm erg}\, {\rm s}^{-1}\,  {\rm cm}^{-3}  \,.              
$$
Note that this equation can be related to the electron density and 
temperature via Eq.~\ref{elredeqt}, $ \Lambda({\rm C}^+) \propto n_e n({\rm C}^+)/\sqrt{T}$. 

To compare the cooling in different directions in the Galaxy, 
it is useful to calculate both the cooling per
neutral hydrogen atom or per nucleon along the line of sight.
We therefore define the cooling per neutral H atom to be:
\begin{equation}\label{ecool1}
l_c \tbond \frac{\int \Lambda({\rm C}^+) ds}{\int n({\rm H}^0) ds}  = 2.89 \times 10^{-20}\frac{N({\rm C}^{+*})}{N({\rm H}^{0})}\, 
	\,{\rm erg}\, {\rm s}^{-1}\,  ({\rm H\,atom})^{-1}\,.
\end{equation}

To estimate  the total hydrogen column density along the sightlines, we
use the ion \ion{S}{2} as a proxy of \ion{H}{1}+\ion{H}{2}, assuming a 
cosmic (meteoric) abundance for sulfur (S/H)$_\odot$ of $10^{-4.80}$ \citep{grevesse98}.
S is not depleted \citep[][and references therein]{savage96}, 
and its second ionization potential of 23.3 eV is high enough to ensure 
that \ion{S}{2} is the dominant ion in both neutral and partially-ionized diffuse gas.
From a study of H$\alpha$ and [\ion{S}{2}] emission lines, 
\citet{haffner99} find that typically \ion{S}{2}/S\,$\sim 0.3$ to 0.8 in the WIM. 
We therefore have  $N($\ion{H}{1}$)+N($\ion{H}{2}$)=({\rm H}/{\rm S})_\odot N($\ion{S}{2}),
and thus we can derive the cooling rate per nucleon as
\begin{equation}\label{ecool2}
L_c = 2.89 \times 10^{-20}\frac{N({\rm C}^{+*})}{N({\rm S}^{+})}  \left(\frac{{\rm S}}{{\rm H}}\right)_\odot \,
	\,{\rm erg}\, {\rm s}^{-1}\,  {\rm nucleon}^{-1}\,.
\end{equation}

Phosphorus is lightly depleted into dust and does not make a good 
proxy of \ion{H}{1}+\ion{H}{2}. But P can be used along with Fe
to study the depletion of the gas (see \S~\ref{multicomp}).

\section{Observations and Data Handling}\label{obs}
\subsection{The Sample}\label{sample}
Our sample is based primarily on a recent {\fuse}\ survey of \ion{O}{6} absorption
toward 100 extragalactic sightlines \citep{wakker03,savage03,sembach03}.
We also searched the {\em HST}/STIS
archive for supplementary or complementary data as well as
for suitable Galactic Halo stars with known distances. 
While many more stars than extragalactic objects were observed with {\fuse}, 
only a few of them are at high enough galactic latitudes to avoid strong
Galactic Disk H$_2$ contamination and have clean-enough stellar spectra for
our purposes (i.e. stars with a simple stellar continuum: metal-poor stars or stars
with rotationally-broadened stellar lines). 
The other criteria for choosing our sightlines are as follows: (1) sightlines with
high-positive velocity \ion{H}{1} resulting in the \ion{C}{2} line overlapping
the \ion{C}{2*} line were rejected; (2) sightlines with strong H$_2$ lines 
that are blended with \ion{C}{2*} $\lambda$1037 were rejected; (3) sightlines
in which the background galaxies' intrinsic Ly$\beta$ line interferes with \ion{C}{2*} $\lambda$1037
were also rejected; (4) only
sightlines with a signal-to-noise ratio of at least 8 per 20 \km\ resolution element
were chosen, to allow reliable measurements. 

\begin{figure}[!h]
\begin{center}
\epsscale{1}
\plotone{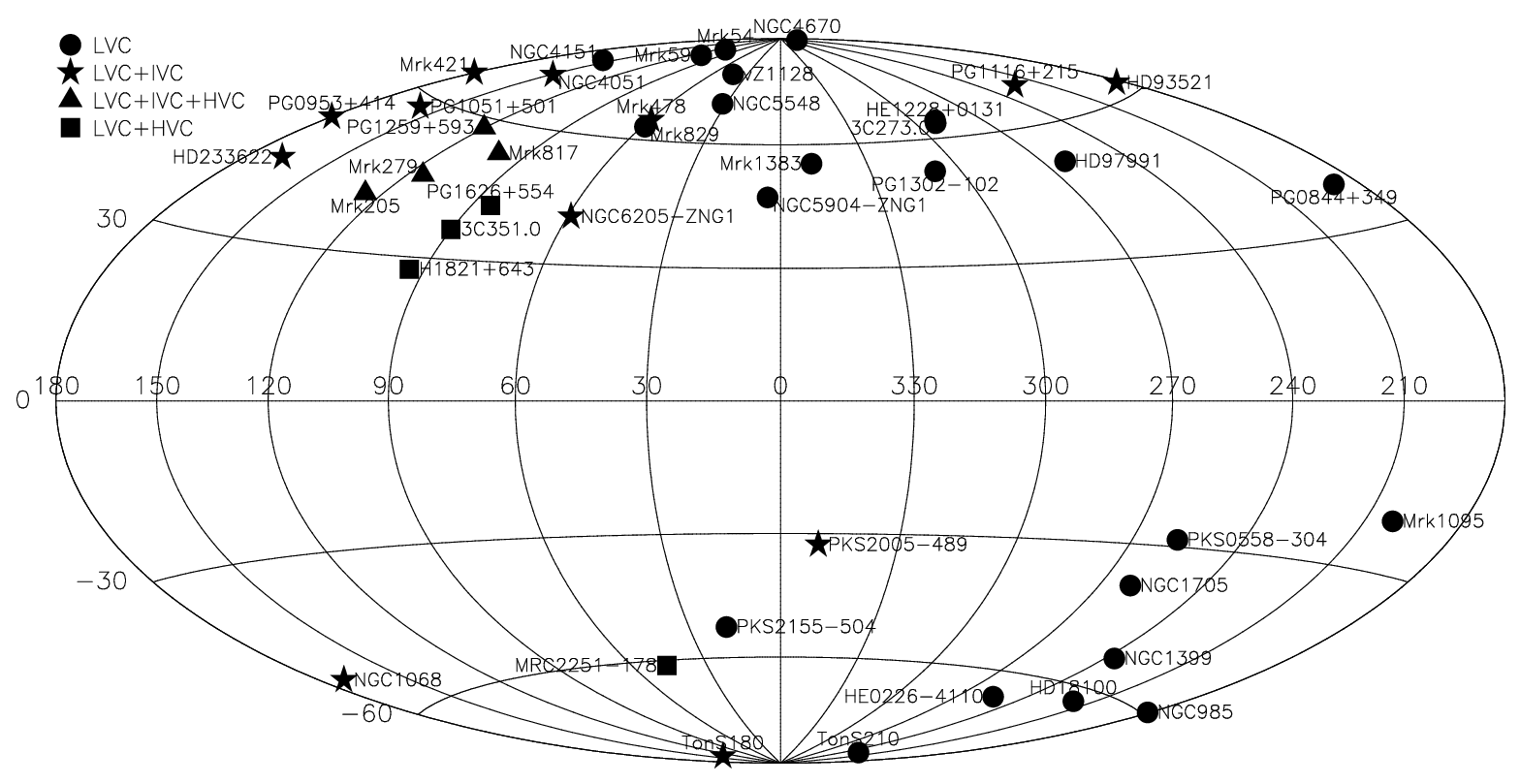}
\caption{Aitoff projection map of the C\,{\sc ii*} survey directions. The Galactic Center is at the 
center and galactic longitude increases to the left. The velocity structure of our sightlines
is complex and we separated them into a low-velocity component (LVC, $\mid v\mid \le 50$ \km), 
an intermediate-velocity component (IVC, $\mid v \mid \ge 50$ \km\ to $ \le 90$ \km),
and a high-velocity component (HVC, $\mid v \mid \ge 90$ \km). The sightlines can have LVCs, IVCs, 
and/or HVCs, as indicated in the figure legend. 
}
\label{map}
\end{center}
\end{figure}

Our final sample consists of 43 objects: 35 QSOs/AGNs, 4 early-type stars, 3 post-AGB 
stars in globular cluster, and 1 subdwarf star. Tables~\ref{t1} and \ref{t1a} list
the principal properties of the extragalactic and stellar objects, respectively. 
The distribution of the targets  on the Galactic sky is shown in Fig.~\ref{map}.
Thirty objects are at $b > +27\degr$, 13
at $b < -21 \degr$. All but three of our targets are situated at 
latitudes $\mid\! b \! \mid > 30 \degr$ because of the large extinction (and large H$_2$ 
column density) in the galactic plane at lower latitudes. 

\subsection{The \ion{H}{1} Emission Spectra}\label{h1s}
We use high spectral resolution ($\sim 1$ \km) \ion{H}{1} 21-cm emission
data for two reasons: (i) to get the component structure along a given 
a sightline; (ii) to measure \ion{H}{1} column densities to compare to 
the other ions.

The \ion{H}{1} column densities were derived by integrating the brightness temperature
over the velocity ranges of the clouds, in \ion{H}{1} observations pointed toward or
near the direction of the background targets, following Wakker et al.\ (2001).
The last column of Tables 1 and 2 lists the source of the \ion{H}{1} data that we
used: the Leiden-Dwingeloo Survey (LDS; Hartmann \& Burton 1997; 35\arcmin\
beam), the Villa Elisa telescope (data courtesy R.\ Morras; 34\arcmin\ beam),
the Green Bank 140-ft telescope (data courtesy E.\ Murphy; 21\arcmin\ beam), the
Jodrell Bank telescope (data courtesy R.\ Ryans; 12\arcmin\ beam), and the
Effelsberg telescope (see Wakker et al.\ 2001; 9\arcmin\ beam). Most of these
spectra were previously shown by Wakker et al.\ (2001).

The major advantage of using the \ion{H}{1} emission line over the Ly\,$\alpha$
absorption lines is that the different clouds in the line of sight can be
separated. The disadvantage is that the beam of the radio telescope is large,
and thus only represents the average column density over a region near the
target. If there is much small-scale structure, the average may differ
substantially from the value in the precise direction toward the target. Wakker
et al.\ (2001) studied this effect, and found that for HVCs $N($\ion{H}{1}) measured with
a half-degree radio beam can differ by up to a factor 2--3 (either way) from the
value measured with a 10\arcmin\ or 1\arcmin\ beam. The distribution of
$N($\ion{H}{1};36\arcmin)/$N($\ion{H}{1};9\arcmin) has a dispersion of about a factor 1.5, or 0.17
dex. Comparing $N($\ion{H}{1}) measured with a 9\arcmin\ beam or with a 1\arcmin\ beam
or through Ly\,$\alpha$ absorption gives a narrower range, corresponding to a
factor up to $\sim$1.25, with a dispersion in the ratio of about a factor 1.15
(0.06 dex). This effect make the value of $N($\ion{H}{1}) in the directions toward the
background targets much more uncertain than would be expected from the noise in
the \ion{H}{1} data. To account for this, we use errors of 0.17 dex for LDS and VE
data, 0.10 dex for Green Bank data, and 0.06 dex for Jodrell Bank and Effelsberg
data, rather than the statistical errors calculated from the noise in the \ion{H}{1} spectra.

In two cases (HD\,18100 and HD\,97991), we instead used 
the column density found from the Ly\,$\alpha$ absorption line
because the FUV observations did not resolve the different  \ion{H}{1}
clouds and the measurements of \ion{H}{1} Ly\,$\alpha$ have smaller errors,
because beam smearing is no longer a problem. 
We also note that in the few cases where there is both 
a  Ly\,$\alpha$ \ion{H}{1} column density and a 21-cm \ion{H}{1} emission column density,
the two measures are consistent (for example, vZ\,1128). 
For sightlines where stars are used to provide the background continuum,
another error could arise because the exact location of the star
relative to the interstellar clouds is unknown. For the \ion{H}{1} emission line 
the column density is derived between us and ``infinity", while for the absorption
line the column density is derived from us to the star. However, because
the stars are at high Galactic latitudes and distant,
this error should remain small. 

The \ion{H}{1} emission spectra are presented in Figs.~\ref{fig2} and \ref{fig2a}, where
the vertical dotted lines indicate the different components obtained 
from the Gaussian profile fitting of the emission line. We also show in this
figure the component number and its associated velocity centroid
and \ion{H}{1} column density in units of $10^{18}$ cm$^{-2}$. The component
numbers appear in the last columns of Tables~\ref{t2} and \ref{t3} that summarize
our measurements. 

\subsection{{\em HST}/STIS E140M Data Reduction}
Several sightlines (see last column of Tables~\ref{t1} and \ref{t1a}) 
were observed with the 
E140M echelle mode of STIS (FUV, $1150-1730$ \AA).
The absorption lines used in this wavelength region are \ion{C}{2*} $\lambda$1335 and 
\ion{S}{2} $\lambda$$\lambda$1250, 1253, and 1259.
The typical spectral resolution of these data is $\sim 7$ \km.
We used the full STIS 3.5 \km\ pixel sampling to analyse the data, 
except if the the signal-to-noise (S/N) ratio was less than 6
per  $\sim 7.0$ \km\ spectral resolution, we binned the data by 
2 pixels. 

Data were reduced within IRAF,\footnote{IRAF is distributed 
by the National Optical Astronomy Observatory which
is operated by the  Association of Universities for Research in Astronomy, 
Inc. under cooperative agreement with the National Science Foundation.}
using the {\sc stsdas} packages. Standard  calibration and extraction
procedures were employed using the {\sc calstis stsdas} version 2.2 routine.

The STIS observations were obtained in the heliocentric frame.  For STIS spectra, 
the absolute wavelength calibration is accurate, and we thus corrected
these spectra to the
Local Standard of Rest (LSR) frame. Once the correction was applied,
a good alignment was observed between the STIS and \ion{H}{1} emission spectra,
and no further correction was deemed necessary (see Figs.~\ref{fig2} and \ref{fig2a}).  
However, the STIS resolution is less than  the spectral 
resolution of the \ion{H}{1} emission data ($\sim$1 \km). Hence, the 
measurements for several
\ion{H}{1} clouds were often combined when 
comparing to the STIS data. For example, in the spectra of HE\,1228+0131
(see Fig.~\ref{fig2}) the first 3 components are combined into one
to compare to the UV absorption at about $-8$ \km, but for component 4
at $+27$ \km\ only one component is observed in both the UV and radio 
spectra. The last column(s) in Tables~\ref{t2} and \ref{t3} indicates 
the \ion{H}{1} components used to compare to the \ion{C}{2*} absorption line 
\citep[the numbers are listed on the lower panels of Figs.~\ref{fig2} and \ref{fig2a} and
follow the cloud definition of][]{wakker01}.

\subsection{{\fuse}\/ Data Reduction}
The {\fuse}\/ instrument consists of four channels: two optimized for the short
wavelengths (SiC\,1 and SiC\,2; 905--1100 \AA) and two optimized 
for longer wavelengths (LiF\,1 and LiF\,2; 1000--1187 \AA). There is,
however, overlap between the different channels, and, generally,
a transition appears in at least two different
channels. For example, \ion{C}{2*} $\lambda$1037 and  \ion{Fe}{2}  
$\lambda$$\lambda$1055, 1063 appear in LiF\,1A, 
LiF\,2B, and SiC\,1A. At $\lambda > 1086$ \AA, \ion{Fe}{2} and \ion{P}{2}  
can be observed in LiF\,2A and LiF\,1B. Note, however, that we mainly 
make use of LiF\,1A for $\lambda = 1000$--1086 \AA\ because of
the better signal and spectral resolution near the 
\ion{C}{2*} $\lambda$1037 line in this channel. The other channels are used to check if 
there are spurious instrumental features.  More complete descriptions of the 
design and performance of the {\fuse}\/ spectrograph are given 
by  Moos et al. (2000) and Sahnow et al. (2000).
To maintain optimal spectral resolution the individual channels 
were not added together. 

Standard processing with version 2.1.6 or higher of the calibration
pipeline software was used to extract
and calibrate the spectra. The software screened the data
for valid photon events, removed burst events, corrected
for geometrical distortions, spectral motions, satellite orbital
motions, and detector background noise, and finally applied
flux and wavelength calibrations.
The extracted spectra associated with the separate exposures 
were aligned by cross-correlating the positions of
interstellar absorption lines, and then combined. 
The combined spectra were finally rebinned by 4 pixels 
($27$ m\AA\ or 7.8 \km\ at 1037 \AA) since the extracted data are oversampled.
This provides approximately three samples per 20 \km\ resolution element.

The {\fuse}\  instrument does not provide an
accurate absolute wavelength scale. It can be inaccurate by about
20 \km. However, with the latest version of pipeline, the relative
wavelength scale remains accurate to better than about 5 \km\ within a  
segment, as found by comparing the measured velocities of many 
interstellar lines within each segment.  To compare \ion{C}{2*} or \ion{P}{2}
to \ion{H}{1}, we employed an approach for adjusting 
absorption-line wavelengths and velocities similar to the one described 
in \citet{wakker03}, and one should refer to their paper
for a more complete description of these issues. To summarize, we used the \ion{C}{2*}
absorption line along with \ion{Si}{2} at 1020.699 \AA\ and 
\ion{Ar}{1} at 1048.220 and 1066.660 \AA\ to compare to the \ion{H}{1}
emission spectra and hence determine the velocity shifts. These 
lines are not strongly saturated, so that usually the deepest absorption
corresponds to the strongest \ion{H}{1} component. We also compare
the {\fuse}\ absorption profiles with STIS absorption profiles whenever
possible to determine reliable velocities.  The spectral resolution of {\fuse}\ 
is far less than that of the \ion{H}{1} 
emission data. Therefore, several \ion{H}{1} clouds were often 
grouped together to compare to the UV absorption lines.

\section{Analysis of the UV Spectra}\label{anal}
In \S~\ref{elecdens}, we showed that \ion{C}{2*} cooling rate  
in erg\,s$^{-1}$\, per H atom is
proportional to $N($\ion{C}{2*})$/N($\ion{H}{1}), 
and the cooling rate in erg\,s$^{-1}$\, per nucleon is
proportional to $N($\ion{C}{2*})$/N($\ion{S}{2}).
The electron density is also directly related to these quantities (\S~\ref{elecdens}). 
Hence, it is essential to have a reliable estimate of the 
column densities of \ion{C}{2*}, \ion{S}{2}, and \ion{H}{1}. 
We already discussed in \S~\ref{h1s} how the \ion{H}{1}  21-cm emission lines
were used to align the velocities of the cloud components and to determine the 
\ion{H}{1} column density. To appreciate the level of saturation of the
absorption lines, we principally use a curve-of-growth method, using 
the \ion{Fe}{2} lines, for which several transitions exist in
the FUV range. Before making any measurements,
we first need to investigate the possible interference of other lines 
with the absorption lines of interest in our work. 

\subsection{UV Absorption-Line Blending}\label{inference}
Galactic components 
(including high-velocity clouds) as well as
extragalactic absorbers can blend with 
the \ion{C}{2*}  absorption lines at 1037.018 and
1335.708 \AA. The last column of Table~\ref{t1} indicates 
which instruments provided measurements of
the different lines. When both instruments
provided measurements, for \ion{C}{2*} the comparison of the two transitions 
allows us to know directly if a line is contaminated.
For the final result we only kept the best-quality 
data (usually from STIS, except if the S/N ratio was not 
good enough). We rejected any sightlines showing serious blending by 
IGM absorption or high-velocity \ion{C}{2}. 

The major contaminant of the \ion{C}{2*} $\lambda$1037 line
is the Lyman 5--0 R(1) line of H$_2$ at 1037.149 \AA, which is at
$+38$ \km\ relative to \ion{C}{2*}. Because 
our objects are at high Galactic latitude and because of our selection
criteria, the amount of H$_2$ is usually small, so that 
\ion{C}{2*} and H$_2$ can be generally easily 
differentiated, as illustrated in Figs.~\ref{fig2} and \ref{fig2a}.  
In all cases where H$_2$ was 
present, we estimated the strength of the $\lambda$1037.149  transition, and 
corrected for it as follows. 
First, we measured two other lines of H$_2$ in the same $J = 1$ rotational
level, including the 7--0 R(1) and 4--0 R(1) lines at  1013.435
and 1049.960 \AA. We chose these lines because
they are in a blend-free part of the spectrum and their strengths
are similar to that of  the 5--0 R(1) line: if 
$f_1=f$(5--0 R(1)) and $\lambda_1= 1037.149 $ \AA, $f_2 = f$(7--0 R(1))
and $\lambda_2=1013.435 $ \AA\ or $f_2 = f$(4--0 R(1))
and $\lambda_2=1049.960 $ \AA\ then $f_1 \lambda_1/f_2 \lambda_2$ 
is 0.91 or 1.15. The close match in $f\lambda$ reduces any
serious saturation effects and the use of two template lines allows us to check 
if these suffer from blending with other features. The 
template lines also reside on the same {\fuse}\/ detector, which 
minimizes differences in the line-spread function. A 
Gaussian absorption function was fitted to the template lines,
which were then 
shifted to 1037.149 \AA, the wavelength of the Lyman 5--0 R(1) line.
Figs.~\ref{fig2} and \ref{fig2a} show the scaled template H$_2$ line on the top of the 
observed $\lambda$1037.149  H$_2$ line. 
We estimated the column density of \ion{C}{2*} $\lambda$1037 absorption 
with and without correcting for the H$_2$ line. In the second case, we 
simply estimated the velocity range over which the \ion{C}{2*} absorption feature
should be present. Both methods generally gave column densities in agreement to
within $\pm 0.05$--0.10 dex, implying that the H$_2$ contamination is negligible.
This is mostly because the \ion{C}{2*} $\lambda$1037 absorption 
does extend much beyond $+20$ \km, while the H$_2$ line gives absorption
centered at $+38$ \km\ on the \ion{C}{2*} velocity scale.

\ion{C}{2*} $\lambda$1335 is generally less likely 
to be blended with other features. For this reason and the
fact that the STIS observations of this line have a higher spectral resolution, we 
favor the STIS observation whenever it has reasonable S/N.

The other ions under consideration are  \ion{S}{2}, \ion{P}{2}, and \ion{Fe}{2}. 
The ion \ion{S}{2} has three transitions in the STIS wavelength range
at 1250.584 \AA\ ($f =5.43 \times 10^{-3} $), 1253.811 \AA\ 
($f =1.09 \times 10^{-2} $), and 1259.519 \AA\ ($f =1.66 \times 10^{-2} $). 
Any possible contamination by IGM absorption can be easily recognized by intercomparing
the three transitions. 

The \ion{P}{2} ion has one useful transition in the {\fuse}
bandpass at 1152.818 \AA\ ($f =2.45 \times 10^{-1} $).
\ion{P}{2} $\lambda$1152 absorption can be contaminated with the \ion{O}{1} 
$^1$D--$^1$D$^0$ $\lambda$1152.151
airglow emission line on the blue side. This airglow line can
cause a continuum-placement problem; yet it is usually minimal for most of the
observations. 

\ion{Fe}{2} has many transitions 
in the {\fuse}\ wavelength range (the \ion{Fe}{2} lines used in this work include 
1055.262, 1063.176, 1096.877, 1112.048,
1121.975, 1125.448, 1127.098, 1133.665, 1142.366, 1143.226, and 1144.938 \AA; 
we used the oscillator strengths derived by Howk et al. 2000).  

Finally, for the stellar sample, photospheric lines can also blend with 
any of the interstellar lines studied here. However, the selected stars 
are either early-type main sequence stars with large projected rotational 
velocity or evolved metal-poor stars, so that any stellar contamination 
is minimized. A comparison between the stellar sample with 
the extragalactic sample does not suggest any systematic effects.

\subsection{Determining Column Densities of \ion{C}{2*}, \ion{S}{2}, \ion{P}{2}, and \ion{Fe}{2}}

The interstellar features of \ion{C}{2*}, \ion{S}{2}, 
\ion{P}{2}, and \ion{Fe}{2} were normalized
by fitting Legendre polynomials to the adjacent stellar or AGN/QSO continuum. 
We present the adopted continuum for the  \ion{C}{2*} measurements for 
each sightline in Figs.~\ref{fig2} and \ref{fig2a}. We note that
the continuum placement is generally simpler in the AGN/QSO spectra
than in the stellar spectra. For example the stellar continuum 
of PG\,1051+501 is complicated by the presence of several stellar 
lines in the \ion{C}{2*} region. For this star, the 
results are uncertain because there is a possibility that the continuum 
may be higher than the one presented in Fig.~\ref{fig2a}.

Since we want to compare column densities of the UV absorption lines 
with \ion{H}{1}, the first steps are to decide which clouds
observed in the UV spectra correspond to which clouds observed in the
\ion{H}{1} spectra, and to decide upon the appropriate absorption-line 
velocity-integration range. The  alignment procedure is discussed in 
\S~\ref{obs}  and the results are presented in Figs.~\ref{fig2} and 
\ref{fig2a}. Usually, the FUV absorption
lines do not resolve the different \ion{H}{1} cloud components. For example,
toward 3C\,273, only two components are clearly separated in the 
STIS spectra, while 4 clouds are found in the \ion{H}{1}
emission spectra. 
Note that for Mrk\,478, Fig.~\ref{fig2}
seems to indicate that \ion{C}{2*} should be separated in two clouds
corresponding to \ion{H}{1} component 2+3 and component 4. But the S/N is low and 
no separation in the \ion{Si}{2} and \ion{Fe}{2} absorption lines is 
observed, so components 2, 3, and 4 were grouped together. 
As discussed in \S~\ref{inference} and presented in Figs.~\ref{fig2} 
and \ref{fig2a}, \ion{C}{2*} $\lambda$1037 is contaminated
by H$_2$ on the red side. We use other absorption lines (\ion{Ar}{1}, \ion{Si}{2},
and \ion{Fe}{2}) to estimate
the velocity range over which \ion{C}{2*} absorption should be measured.
The velocity ranges over which the profiles were integrated are indicated
in Figs.~\ref{fig2} and \ref{fig2a}. 
Depending on the strength of H$_2$, a comparison of the measured values 
with and without removing H$_2$ (and thus usually measuring
over a slightly larger velocity range) gave column densities that were consistent 
within less than 0.05--0.10 dex. An exception is Mrk\,1095,
for which the H$_2$ absorption line 
is particularly strong and broad, and where the difference between
estimating the \ion{C}{2*} column density with and without removing the
H$_2$ line was about 0.3 dex. In summary, a combination of the 
\ion{H}{1} emission spectra and  uncontaminated FUV absorption lines of various metals
were used to estimate the velocity range over which we measured the 
equivalent widths and column densities of \ion{C}{2*}. 

The adopted uncertainties for the derived 
column densities and Doppler parameters are  $\pm 1\sigma$.  
These errors include the effects of statistical noise, fixed-pattern
noise, and the systematic uncertainties of the
continuum placement, the H$_2$ contamination, and the  velocity range over 
which the interstellar absorption lines were integrated.  

To calculate the column densities two 
methods were used: the curve of growth and the apparent optical depth 
methods. A curve of growth (COG) for the \ion{Fe}{2} 
and the \ion{S}{2} lines was constructed independently from the
measured equivalent widths of these species. 
A single-component Gaussian COG
was constructed in which the Doppler parameter $b$ and 
the column density $N$ were varied to minimize the $\chi^2$ between the observed equivalent
widths and a COG model. The resulting column density and 
Doppler parameter are given for each sightline in Table~\ref{t2} in 
columns 2 and 3 for \ion{Fe}{2}. The Doppler parameter for \ion{S}{2}
is given in column 8. It generally is consistent to within the $1\sigma$
errors with the $b$-values derived from the \ion{Fe}{2} lines; hence we
also derived the column density of \ion{S}{2} using the $b$-value
from \ion{Fe}{2}. The latter has a smaller error because there are many 
more \ion{Fe}{2} absorption lines (between 5 and 11) than \ion{S}{2} absorption lines (a maximum
of 3 can be measured). 

We further measured the column density 
of the \ion{S}{2} lines using the apparent optical depth method. In this method, 
the absorption profiles are converted into the apparent
optical depth (AOD) per unit velocity, $\tau_a(v) = \ln[I_{\rm c}/I_{\rm obs}(v)]$, 
where $I_{\rm obs}$ and $I_{\rm c}$ are the observed intensity and the
estimated continuum intensity, respectively.  $\tau_a(v)$ is related
to the apparent column density per unit velocity, $N_a(v)$  through the relation 
$ N_a(v) = 3.768 \times 10^{14} \tau_a(v)/[f \lambda(\rm \AA)]$ 
cm$^{-2}$ (\km)$^{-1}$ \citep{savage91}.
The integrated apparent column density is equivalent to the 
true integrated column density in cases where 
the lines are resolved. The results for the apparent column density method generally compared
favorably to the column density derived from the COG method, except when the lines
were saturated (in that case $ N_a$ is a lower limit, lower than $N$ derived from 
the COG). The adopted column densities of \ion{S}{2} are given in column 7
of Table~\ref{t2}. If no saturation effect was observed, the apparent column 
density was adopted. Otherwise the \ion{S}{2} column density from the COG (using the 
$b$-values derived with the \ion{Fe}{2} absorption lines) was adopted, except
in three cases (H\,1821+643, NGC\,4151, and PG\,0953+414) where the $b($\ion{S}{2})
is significantly smaller by about 4--5 \km\ than $b($\ion{Fe}{2}). In those 
cases, we adopted $b($\ion{S}{2}) to obtain $N($\ion{S}{2}).

For the \ion{C}{2*} and \ion{P}{2} absorption lines, the 
equivalent widths were used to deduce the column density using 
the COG derived from the \ion{Fe}{2} and \ion{S}{2} lines. 
We measured the apparent column density of \ion{C}{2*}
and \ion{P}{2} following the same method as that for \ion{S}{2}. 
For \ion{C}{2*}, the apparent column density
is presented in column 4 of Table~\ref{t2}. Column 5 gives
the adopted column density of \ion{C}{2*}. If the 
numbers in columns 4 and 5 are the same, the apparent optical
column density was adopted. Otherwise,
the column density derived using the COG was adopted, but one
can notice that saturation effects are generally small. The 
COG of \ion{Fe}{2} was generally adopted to determine the column 
density of \ion{C}{2*}, except in cases where
we were not able to derive it (no {\fuse}\ data or the {\fuse}\ data
did not separate the cloud components, and therefore
$b($\ion{S}{2}) was used) and in the case of H\,1821+643 (component 3+4+5), where we 
adopted the COG of \ion{S}{2} because the column density derived
from the COG of \ion{Fe}{2} was too small compared to the apparent
column density (i.e. $b($\ion{Fe}{2}) is too large). 

Because P has a low cosmic abundance, the available transition 
of \ion{P}{2} did not generally suffer from any saturation;
hence the apparent column density was adopted. 

We note that the good agreement between the AOD and the COG
column densities ensures that the COG parameters derived for \ion{Fe}{2}
are similar to those of \ion{C}{2*}. The only case where we found
this not to be the case was toward H\,1821+643, where  $b($\ion{Fe}{2}) appears 
to be too large. Thus, the different observed ions are probably 
formed in regions with similar physical conditions. 

\section{Summary of the Properties of the Gas Studied}\label{multicomp}
Before presenting and discussing the observed cooling
rates, we review several properties of the clouds
being studied: their  ionization
structure, temperature, dust content, and metallicity. 
These affect the interpretation of the absorption-line
results. 

The velocity structure of our sightlines
is complex, containing both disk gas and halo gas, the
latter in the form of intermediate- and high-velocity
clouds (IVCs, HVCs). We separate them for the rest of 
this work into low-velocity clouds (LVCs), defined 
as gas moving at velocities compatible with a simple model 
of differential galactic rotation, and IVCs and HVCs, defined
as gas moving at velocities larger than those predicted 
by simple models of differential galactic rotation \citep[e.g.,][]{wakker01a}.
For the IVCs and HVCs, we use the nomenclature of \citet{wakker91},
\citet{kuntz96}, and \citet{wakker01a}. Typically,  
HVCs have absolute LSR velocities greater than 90 \km, 
LVCs have absolute LSR velocities smaller than 50 \km, and IVCs 
have LSR velocities between 40 \km\ and 90 \km.

\subsection{Low-Velocity Clouds}\label{lvcintro}
Usually, multiple low-velocity \ion{H}{1} components are  
unresolved in the UV absorption lines (see Figs.~\ref{fig2} and 
\ref{fig2a}, and \S\S~\ref{obs} and \ref{anal}). 
Such a complex structure makes it more difficult to 
interpret the column densities and their ratios, 
as the absorptions from different clouds can potentially 
blend together. These different clouds may have different
ionization structure. Further, even in one cloud different
ions may originate in different parts of it, some in 
the WNM, some in the WIM.

Toward HD\,93521 \citet{spitzer93} found  10 interstellar clouds ranging 
in velocity from $-66.3$ to $+7.3$ \km. Nine of these clouds are warm ($T\sim 6000$ K) and
composed of a mixture of ionized and neutral gas. 
The tenth cloud has a very low column density and traces cold gas ($T\sim 500$ K). 
Using their measurements we computed the cooling rates
from Eqs.~\ref{ecool1} and \ref{ecool2}. These are shown 
in Fig.~\ref{hd93} along with our {\fuse}\ measurements.
With {\fuse}\ only the 2 main clouds (LVC and IVC) 
are detected in the \ion{C}{2*} $\lambda$1037 absorption lines,
but the derived cooling rates give a good approximation for the IVC and LVC 
components, even though there is weak H$_2$ contamination. 
This comparison shows that, even though the spectral resolution of
{\fuse}\ does not allow to resolve all the different components, 
we still can derive the correct cooling rates for the 
LVC and IVC. 

A few other sightlines in our sample with high S/N data also have been 
thoroughly analyzed, and we briefly summarize some 
of their physical properties. 
\citet{howk03,howk04}  showed that the gas along the path to vZ\,1128  
contains a large fraction of warm ionized gas ($N($\ion{H}{2}$)/N({\rm H}) = 0.46 \pm 0.02$).
The WIM and the WNM are kinematically associated and hence
closely related. There is no evidence of cold clouds 
toward HD\,18100; the gas is mostly WNM \citep{savage96a}. 
Finally, toward 3C\,273 \citet{savage93} found that the gas is 
warm and partially ionized.

The large full width at half-maximum (FWHM, which is related to the Doppler parameter, $b$, via 
${\rm FWHM} = 2\sqrt{\ln 2} b $) of the 
\ion{H}{1} emission indicates the presence of a warm component 
($T_{\rm kin} \ga 5000$ K; ${\rm FWHM} \ga 15 $ \km), but also of a colder component (a few hundred
K to 3000 K) toward most of our 
sightlines \citep[see the listed FWHMs in][]{wakker03}. These temperatures 
are lower limits, because turbulence is not taken into account. The \ion{H}{1}
column densities are dominated by warmer components. 
Toward a few sightlines, a weak cold component is present
(${\rm FWHM} \sim 3$--4 \km, $T_{\rm kin} \la 300$ K). This occurs, 
for example, for component 2 in the spectrum of NGC\,985 and component
3 in the spectrum of Mrk\,1095. But note that the 
\ion{H}{1} column densities for these two components are less than 10\%
of the warm \ion{H}{1} component. Fig.~\ref{fig2} shows for these 
two sightlines that H$_2$ is relatively abundant. 

The depletion pattern of Fe (and to a lesser extent the depletion of P) 
further shows that warm gas dominates our sightlines.\footnote{We define [X\,{\sc ii}/H\,{\sc i}] as the  
abundance ratio in logarithmic solar units (from 
the solar meteoric values of Grevesse \& Sauval 1998): 
[X\,{\sc ii}/H\,{\sc i}$] \tbond \log(N($X\,{\sc ii}$)/N($H\,{\sc i}$)) - \log({\rm X/H})_\odot$.
We also use throughout this paper the definition of depletion as the deficiency
of an element in the gas-phase because of the incorporation of the
element into dust grains.}
\citet{savage96} and references therein 
\citep[but see also][]{welty99,wakker00,jenkins03} show
a general progression of increasingly severe depletion from warm halo 
clouds, to warm disk clouds, to colder disk clouds. In the
halo, a ``typical'' value of [\ion{Fe}{2}/\ion{H}{1}]  is $-0.6$ dex, in the warm disk
$-1.5$ dex, and in the cold disk $-2.2$ dex. 
Fig.~\ref{depl} and Table~\ref{t3} show that for LVCs [\ion{Fe}{2}/\ion{H}{1}] 
is between $-0.5$ and $-1.4$ dex, implying that the LVC components
mostly trace warm gas. 
There is a clear dependence between [\ion{Fe}{2}/\ion{H}{1}] or 
[\ion{P}{2}/\ion{H}{1}]  and \ion{H}{1}. However, in this diagram, species
that can live in both neutral and ionized regions (\ion{P}{2} and \ion{Fe}{2})
are compared to \ion{H}{1} (tracing only neutral gas). This dependence
is therefore difficult to interpret because both ionization and depletion 
play a role in the observed distribution of [\ion{P}{2}/\ion{H}{1}] 
and [\ion{Fe}{2}/\ion{H}{1}]. The 
middle diagram of Fig.~\ref{depl} (but see also Table~\ref{t3} 
for the uncertain measures and lower limits
that are not included in Fig.~\ref{depl}) shows that within the errors 
S is not depleted, and therefore is a good proxy for \ion{H}{1}+\ion{H}{2}. But it 
also shows several high values of [\ion{S}{2}/\ion{H}{1}] 
even for \ion{H}{1} column density of about $10^{20}$ cm$^{-2}$, 
implying that ionization corrections would be necessary to really
understand the depletion of Fe and P at these column densities.

If \ion{H}{2}/\ion{H}{1} is low, [\ion{S}{2}/\ion{H}{1}] is expected to be 
about solar, while for more highly-ionized cloud
[\ion{S}{2}/\ion{H}{1}] should be supersolar. For 6/19 sightlines [\ion{S}{2}/\ion{H}{1}]
is supersolar by at least 0.1--0.2 dex. In several cases, the errors 
are asymmetric with the upper error bar larger than the lower error 
bar, implying that $N($\ion{S}{2}) could be larger and hence ionization could 
be significant.

We can actually directly estimate the ionized fraction toward   NGC\,5904-ZNG1 (M\,5) and
NGC\,6205-ZNG1 (M\,13) via the pulsar dispersion measures (DM\,$= N_e = \int n_e ds $). 
\citet{reynolds91} found DM\,$=29.5$ and 30.5 cm$^{-3}$\,pc toward M\,5 and M\,13,
respectively, implying $\log N_e = 19.96$ and 19.97 dex. Under the assumption 
that all the He is neutral, $N($\ion{H}{2}$) = N_e  $, so $N($\ion{H}{2}$)/N({\rm H}) = 0.22$
toward M\,5 and 0.40 toward M\,13. Even if $N($\ion{H}{2}$) \sim 0.8 N_e  $ \citep[i.e. taking
into account that some electrons come from singly- and doubly-ionized helium,][]{howk04},
there is still a significant fraction of ionized hydrogen along these sightlines.
Ionization may be substantial along many of our sightlines, 
(see \S\S~\ref{elecdens} and \ref{lvcdiscuss}).
This is not  unexpected, since our sightlines go
directly through the WIM revealed by 
H$\alpha$ emission at high galactic latitudes. The low
density WIM fills more than 20\% of the volume 
within a 2 kpc thick layer around the midplane and
has $n($\ion{H}{2}$)/n($\ion{H}{1}$) > 15$ at $T = 8000$ K
\citep{reynolds93}.   We also note that along all these sightlines
\ion{O}{6} absorption  was detected. While the relationship
between the highly- and weakly-ionized gas is not well 
known, radiation from the hot gas is an important source of 
photoionization \citep{slavin00}.

Since we selected sightlines with low H$_2$ column densities, we should also
not expect much dust to be present. And since most of the dust
is in cold clouds, we should find mostly warm gas in our sightlines. 
These expectations are indeed borne out by the results
discussed above, and mostly warm gas is traced. 
This gas is a mixture of neutral and ionized gas. While the importance
of ionized gas was only demonstrated directly toward a few sightlines 
in our sample, we believe that the ionized fraction is substantial
toward most of our sightlines.

\subsection{Intermediate-Velocity Clouds}
IVCs have absolute LSR velocities between $\sim$40 and 90 \km. 
Their lower velocities compared to HVCs can make the absorption profiles blend
with lower-velocity gas which happens 
in two cases (Mrk\,59 and NGC\,4151, see Tables~\ref{t2} and \ref{t3}).

The main IVC surveyed is the intermediate-velocity arch (IV Arch). 
Upper limits or measurements of the cooling rate were 
obtained toward several regions of this complex, labeled 
in the last column of Table~\ref{t2} (IV Arch, IV5, IV9, IV16, IV18, IV26, and
LLIV; see Wakker 2001 for a complete description of these different regions) and
see  Figs.~\ref{fig2} and \ref{fig2a} for  the LSR velocities of these features. 
The IV Arch lies in the general direction of HVC Complex C, but the 
Low-Latitude Intermediate-Velocity Arch (LLIV), as its name indicates, is 
at lower latitude. The  IV Arch lies at
a $z$-height between 0.5 and 3 kpc \citep{wakker01a}. The LLIV may 
be a high-$z$ interarm region. 

Previous studies show that the  metallicity of the IV Arch  
is essentially solar \citep[see][and references therein; see also \S~\ref{ivcdist}]{wakker01a}. 
Fig.~\ref{depl} confirms that [\ion{S}{2}/\ion{H}{1}] is essentially solar, although 
some high values for IV16 (PG\,0953+414) and LLIV (Mrk\,205) imply that the gas
may be ionized.  The depletion
for P is small, but varies for Fe between $-0.1$ dex and
$-0.9$ dex. The changes in ionized-fraction and depletion indicate
that the physical conditions must vary in the IV Arch, but
the low depletion implies warm clouds. There is also
hot gas detected via \ion{O}{6} absorption \citep{savage03}, though
no clear-cut association between the \ion{H}{1} and \ion{O}{6} is found. 

A component of the IV Spur (S1) is detected toward PG\,1116+215 and is 
an extension of the IV Arch. \citet{kuntz96} derived a distance 
bracket of 0.3--2.1 kpc for the IV Spur. It has a solar metallicity and the depletion
of Fe is small (see Table~\ref{t3}). The ratios \ion{S}{3}/\ion{S}{2}
and \ion{Fe}{3}/\ion{Fe}{2}  are about 0.2, implying the presence
of ionized gas.  \ion{O}{6} absorption was observed
along this sightline at velocities similar to those for the
weakly-ionized species \citep{savage03}, indicating the presence
of kinematically hot gas associated with the warm partially-ionized gas. 

Toward NGC\,1068, an unclassified  IVC is detected at $-53$ \km. Toward  
PKS\,2005--489 and Ton\,S180, unclassified positive-velocity IVCs are detected 
at $+75$ \km\  and $+40$ \km, respectively, (see Fig.~\ref{fig2}). 
\ion{O}{6} absorption lines are observed in
the FUV spectra of these objects \citep{wakker03,savage03}, showing
as well for these IVCs the presence of a hot phase kinematically associated
with the warm gas for these IVCs. 

Toward Mrk\,478 and NGC\,6205-ZNG1, a $3 \sigma$ upper limit on the 
\ion{C}{2*} column density was estimated for Complex K (although not 
very stringent toward Mrk\,478 because of the low signal-to-noise of this spectrum). 
Complex K lies near the direction of Complex
C, and is defined as having LSR velocity between $-95$ and $-60$ \km.
Its properties and origin are not well known \citep{wakker01a}, but
faint H$\alpha$ emission was detected by \citet{haffner01},
and \citet{savage03} also detected \ion{O}{6} absorption. 
Therefore, ionized and highly-ionized gas are also present in this complex. 

Toward MRC\,2251--178 a $3 \sigma$ upper limit was estimated 
for the IVC component that might be considered 
part of the Complex gp, although this limit is not 
very stringent because of the low S/N ratio. 
This complex  probably has solar
metallicity \citep{wakker01a}.

In summary, several IVCs are probed, but their dominant gas-phase
is warm with both the presence of ionized and neutral gas. They
have also a solar metallicity.

\subsection{High-Velocity Clouds}\label{hvcgeneral}
HVCs have absolute LSR velocities larger than $\sim$90 \km. 
HVCs are well separated from the lower-velocity components, but we
mostly derive $3\sigma$ upper limits of \ion{C}{2*} absorption for the
two HVC complexes that were investigated: Complex C and the Outer Arm (OA). 
A compact HVC (WW84) is also observed in \ion{H}{1} toward Mrk\,205,
with a limit determined for \ion{C}{2*} absorption.
Other HVCs exist along other sightlines included in our survey, 
but no \ion{C}{2*} absorption line limit could be measured because of blending with
other lines (generally  \ion{C}{2} or H$_2$). 

Complex C consists of a large assembly of high-velocity
gas, covering 1600 square degrees of the northern galactic sky, 
between $l\sim 30\degr$ and $150\degr$. It has LSR velocity ranging
between $-90$ and $-200$ \km. We were able to derive upper limits on the
\ion{C}{2*} absorption for several regions of this complex: 3C\,351 (CIB), Mrk\,205 (C-south),
Mrk\,279 (C-south), Mrk\,817 (CIA), and PG\,1626+554
(CI) and one (tentative) measurement toward  PG\,1259+593 (CIIIC).
Complex C has a subsolar metallicity of $\sim 0.14 Z_\odot$ 
\citep{wakker99,gibson01,richter01,collins03,fox04}. 
Toward PG\,1259+593, we found for the HVC component
[\ion{S}{2}/\ion{H}{1}]\,$=-0.71 \pm 0.22$ dex and [\ion{Fe}{2}/\ion{H}{1}]\,$=-0.83 \pm 0.11$ dex 
(the solar meteoric abundances are from Grevesse \& Sauval 1998).
S is not depleted into dust grains, thus its low value reflects a low abundance
of $Z \simeq 0.15 Z_\odot$, consistent with previous results. 
Note that the $3 \sigma$ upper limit for [\ion{P}{2}/\ion{H}{1}$]\le -1.05$ dex suggests a lower
abundance for this element.  We note
that Fe is underabundant by a similar amount as S,
showing that there is little dust in the HVC along this sightline. 
Our value for [\ion{Fe}{2}/\ion{H}{1}] differs from the results of \citet{richter01}
and \citet{collins03} for the reasons given in the footnote to Table~\ref{t2}.
Complex C contains some ionized gas, as revealed
via the detection of \ion{O}{6} and H$\alpha$ \citep{sembach03,tufte98,wakker99},
at similar LSR velocities as observed for the neutral gas \citep{fox04}. 
The distance to Complex C 
is not well known, but it is at least $> 6$ kpc \citep{wakker01a}.
Its exact origin is still debated, but its properties
indicate that the Complex C is most certainly extragalactic. 

The Outer Spiral Arm (OA) is detected toward H\,1821+643 in \ion{H}{1} at $-128$
and $-87$ \km\ at low latitude ($b\sim 27\degr$). 
The velocities are only 20--30 \km\ higher
than expected from galactic rotation curve at galactocentric 
radii of $\sim20$ kpc \citep{wakker01a}. 
The metallicity is essentially solar, although for the lowest 
\ion{H}{1} component number 1, the supersolar [\ion{S}{2}/\ion{H}{1}] implies 
a large fraction of ionized gas. 
There is little or no dust containing Fe for both 
components of the OA (see Fig.~\ref{depl}).
Highly-ionized species were reported toward H\,1821+643 by 
\citet{savage95}, \citet{sembach03}, and \citet{tripp03}.

Toward Mrk\,205, the HVC observed at $-202$ \km\ is
referred to as a very high-velocity compact cloud. 
New {\fuse}\ data (Wakker et al. 2004, in preparation) show
it has a metallicity of $\sim0.1$--0.2 solar. \citet{braun00}
mapped the cloud at 1\arcmin\ resolution
and concluded that it consists of a cold ($T \sim 85$ K) core
embedded in a warm envelope. \citet{braun01} also argue
that the cloud lies at a distance of 300--900 kpc. 

\section{Cooling Rates}\label{coolmeas}
The \ion{C}{2} cooling rates were computed with Eqs.~\ref{ecool1} and \ref{ecool2}
and are summarized in Table~\ref{t3} in erg\,s$^{-1}$\, per H atom
(column 2 using the \ion{H}{1} emission measurement) 
and in erg\,s$^{-1}$\, per nucleon (in column 3 where \ion{S}{2} 
was used as a proxy for H, respectively). 
Note that Table~\ref{t3} differs from Table~\ref{t2} in 
the sense that types of clouds (LVCs, IVCs, or HVCs) are
grouped together. The cooling rates in erg\,s$^{-1}$\, per H atom are also shown
graphically in Fig.~\ref{cool} against the \ion{H}{1} column density. 
This figure not only shows possible systematic trends of 
the observed cooling rates with the amount of \ion{H}{1}, 
but also compares our results with others obtained 
from local FUV observations and IR observations at high Galactic latitudes
and also in damped Ly\,$\alpha$ systems toward QSOs. Those other observations
are discussed below and are summarized in Table~\ref{t5}, along with
our observations.

\subsection{Low-Velocity Clouds}\label{lvcdist}
\subsubsection{Mean and Range of the Cooling Rates}
The \ion{C}{2} cooling rates for the LVCs span about an order of 
magnitude lying between $-26.3$ and $-25.3$ dex (see Table~\ref{t3}, Figs.~\ref{cool} 
and \ref{coolbis}), if we exclude  Mrk\,1095 ($\log  l_c = -26.70 $ dex) and limits or uncertain values
(see below for more details on these sightlines).
The mean cooling rate per H atom is $\log [(\Sigma_1^N l_c)/N] = 
\log \langle l_c \rangle = -25.70 \pm\,^{0.19}_{0.35} $ dex. 
The mean cooling rate per nucleon, using \ion{S}{2}, is 
$\log \langle L_c \rangle = -25.65 \pm\,^{0.11}_{0.15} $ dex.
The errors given here are the deviation around the mean.
The typical $1\sigma$ errors from the measurements are 
small enough to imply that the observed dispersion is a real change of the
cooling rate from sightline to sightline. Note
that the mean value was derived excluding the two sightlines where LVC and IVC components 
are blended (see Table~\ref{t3}). Note also that the sample to obtain $\langle L_c \rangle $  
is much smaller than the sample to obtain $\langle l_c \rangle $ and is biased toward 
\ion{H}{1} column densities larger than 10$^{20}$ cm$^{-2}$ (see Fig.~\ref{depl}), 
where ionization effects are less likely. 
Using the same sample to calculate the cooling rate per H atom  and per nucleon, 
$\log \langle l_c \rangle = -25.61 \pm\,^{0.17}_{0.27} $ dex. 
In Fig.~\ref{s2c2}, we compare the cooling rates per H atom 
and per nucleon. The straight line is a 1:1 relationship. There
is a scatter of about $\pm 0.1$ dex around this line, except for 4 cases
(where $\log N($\ion{H}{1}$) \le 20$ dex) 
that depart by more than $+0.1$ dex.

\subsubsection{Dispersion of the Cooling Rates}

Fig.~\ref{coolbis} shows the LVC cooling rate
against the \ion{H}{1} column density for $\log N($\ion{H}{1}$)> 19.5$ dex. 
The solid line shows the mean cooling rate, with the dotted lines 
giving the $1\sigma$ dispersion around the mean.
Two features are apparent from this figure:
(1) a large scatter of $l_c$ of about 0.5 dex (a factor $\sim$3
change from sightline to sightline) at any given $N($\ion{H}{1}); 
(2) a decrease of $l_c$ with increasing $N($\ion{H}{1}). 

The decrease of the cooling rate per H atom with increasing $N($\ion{H}{1})
can be understood as an ionization effect.  In \S~\ref{lvcdiscuss}, we 
show that a substantial fraction of the observed \ion{C}{2*} may
come from the ionized region of the cloud, where more electrons are present.
In \S~\ref{elecdens} we saw that collisional excitation 
of \ion{C}{2} with electrons is far more efficient than
with hydrogen atoms in the diffuse warm gas. 
So, at  low \ion{H}{1} column density (less than a few $10^{19}$ cm$^{-2}$), 
this effect is very apparent because the fraction of photoionized
gas is significant ($\ga 90$\% if $\log N_e \sim 19.9$ dex, see \S~\ref{lvcintro}).
For example, for $\log N($\ion{H}{1}$)< 19.5$ dex, the largest deviation 
from the mean cooling rate is observed, toward 
HE\,1228+0131 in component 4 at $+27$ \km, where
$\log l_c \approx -24.80$ dex, nearly 8 times
higher than the mean value. The cooling rate derived with \ion{S}{2} 
implies a lower value, $-25.30$ dex, still a factor
$\sim 3$ times higher than the mean value. The ratio
[\ion{S}{2}/\ion{H}{1}$]\sim +0.5$ dex also implies a substantial fraction of ionized gas
along this sightline. A similar high value for the cooling
rate is also found toward 3C\,273, only 0\fdg83 away from HE\,1228+0131,
for the $+25$ \km\ component number 4. 
At higher \ion{H}{1} column densities, the fraction
of ionized gas must diminish to about 10--50\% (assuming 
$\log N_e \sim 19.9$ dex). In Fig.~\ref{coolbis},
the sightline for which the cooling rate departs most from the mean 
is Mrk\,1095. It has the lowest cooling rate, a factor 10 
smaller than $\langle l_c \rangle$ for the largest \ion{H}{1} column density 
in our sample (20.97 dex). If $\log N_e \sim 19.9$ dex, the gas along 
this sightline is almost entirely neutral. The depletion of Fe for this 
sightline informs us that the gas is mostly warm. Models of multi-phase 
gas predict very low \ion{C}{2} cooling rates for the WNM (a factor $\sim 10$ times smaller than the 
cooling rates in the WIM or the CNM, Wolfire et al. 1995) roughly 
on the order of the cooling rate measured toward Mrk\,1095. 

In Fig.~\ref{maplvc}, we show the cooling rates projected
on the Galactic northern (left hand-side) and southern (right hand-side) 
sky, where the  \ion{H}{1} contours show the column density 
for gas with $\mid\!v_{\rm LSR}\!\mid \le 50 $ \km. 
This figure confirms the above discussion: in regions
with higher \ion{H}{1} columns, the \ion{C}{2*} cooling rate in LVCs 
is lower in both the south and north Galactic sky and vice-versa. 
There does not appear to be a latitude or longitude dependence or a
north-south asymmetry within the errors. We also did not find 
any relation between the cooling rate and the distance
for the stellar sightlines given in Table~\ref{t1a}, 
suggesting that the bulk of the observed gas is at $z < 800 $ pc 
(smallest $z$-height in our sample). We note, however, a
decrease in the cooling rates with $z$-height when the LVC,
IVC, and HVC are considered together. For IV Arch at 
$z \sim 1$ kpc we find that on average the cooling is a factor 2 lower
(see \S~\ref{ivcdist}) and for Complex C, at $z > 6$ kpc
it is 20 times lower (see \S~\ref{hvcdist}).

The observed dispersion in the cooling rates in the LVCs certainly 
does not have a single explanation. The change in the ionized
fraction explains why the cooling rates decrease at 
higher \ion{H}{1} columns. In the WNM, a change 
in the temperature of the gas can affect which cooling 
process dominates. For example, \citet{wolfire03} show 
that for $ 500 \la T \la 8000 $ K the cooling due to 
[\ion{O}{1}] 63 $\micron$ becomes more important than 
[\ion{C}{2}] 158 $\micron$, and 
collisional excitation of Ly$\alpha$  dominates the cooling if 
$T \ga 8000$ K. Also, variations in the heating 
that balances the cooling are expected in gas heating 
models \citep{reynolds99,wolfire03}.
To explain the emissivity of \ion{C}{2} in the WNM, 
photoelectric heating from small grains and 
PAHs appears to be the dominant heating process
\citep[e.g.,][]{wolfire03}.
So, the dust-to-gas fraction may also vary, although there is no direct
evidence for this variation from our observations. The cooling 
rate per H atom does not increase with smaller values of [\ion{Fe}{2}/\ion{H}{1}], 
but note that the depletion of elements such as Fe does not really determine 
if the total dust-to-gas fraction is varying, since dust and PAHs are 
believed to be mostly composed of C and Si \citep{draine03}.
Pure photoionization models of the WIM have a heating rate
per unit volume about $\sim 10^{-24} \times n_e^2$, so photoionization of H can provide only
part of the heating of the diffuse gas because $n_e \simeq 0.08$ cm$^{-3}$ 
\citep{reynolds92a,reynolds99,slavin00}. 
Photoelectric grain heating of the WIM can provide a supplemental
heating mechanism \citep{reynolds92a,draine78}. 
However, other sources are possible,
such as the dissipation of interstellar plasma turbulence 
that may even dominate over photoionization in regions 
where $n_e < 0.1$ cm$^{-3}$, because the heating rate per unit volume 
is in that case about $\sim 10^{-25} \times n_e$ \citep{minter97,reynolds99}. 
Other heating processes in the WIM may include
magnetic-reconnection \citep[e.g.,][]{raymond92}, 
or coulomb collisions by cosmic rays \citep[e.g.,][]{skibo96},
although it is not clear if they are a significant source of heating.  

\subsection{Intermediate-Velocity Clouds}\label{ivcdist}
The main IVC investigated is the IV Arch. 
Table~\ref{t3} presents the cooling rate for the individual 
sightlines through this complex. 
The mean cooling rate per nucleon in the IV Arch using 
\ion{S}{2}, $\log \langle L_c \rangle = -25.98 \pm\,^{0.09}_{0.12}$ dex. 
The errors given here are the deviation around the mean.
The mean cooling rate per H atom appears to be
$\log \langle l_c \rangle = -25.76 \pm\,^{0.22}_{0.46}$ dex. 
But much higher values are found toward Mrk\,205 ($-25.45$) and
PG\,0953+414 ($-25.51$). However, 
the IV Arch is substantially ionized toward these 
two sightlines and indeed, if those sightlines are not included or
if the ionization is corrected for, 
the mean cooling rate per H atom would be 
$\log \langle l_c \rangle = -25.98 \pm\,^{0.14}_{0.20}$ dex, in agreement
with the mean derived using \ion{S}{2}. 
The mean cooling rate of $-25.98$ dex in the IV Arch is about two times 
smaller than the mean cooling rate of the LVCs. 

In Fig.~\ref{ivczoom}, we show the cooling rate  against the 
\ion{H}{1} column density for the IV Arch components. 
There may be an increase of the cooling rate with $N($\ion{H}{1}), 
(but the number of data points is small) 
if the low limit toward PG\,1051+501 is not taken into account. 
The stellar continuum of PG\,1051+501 is uncertain, producing
uncertain results. In Fig.~\ref{mapivc}, the cooling rates
of the individual sightlines are overplotted on an \ion{H}{1}
contour map, showing again that the higher values of the cooling rates 
are in the higher \ion{H}{1} column density regions.

The strength of the FUV radiation field that heats 
the gas may be lower in the IV Arch (resulting in a weaker
cooling in the IV Arch) because the IVC is more distant 
than the LVCs.
 The slight increase of the cooling rate with increasing $N($\ion{H}{1}) 
suggests that  the emissivity of \ion{C}{2} is higher in the denser
and neutral regions of the IV Arch. 
This behavior contrasts with the decrease of 
$l_c$ with increasing $N($\ion{H}{1}) observed in the LVCs, 
but this effect could just be because of the small number 
of sightlines.

In the IV Spur (PG\,1116+215, S1 component),
about 20\% of the gas is in ionized form. The cooling rate per nucleon is 
$-25.65$ dex (mean of the cooling using \ion{S}{2}). 
The Spur region is believed to be an extension
of the IV Arch. Including  $\log l_c = -25.65$ dex and 
$\log N($\ion{H}{1}$)= 19.83$ dex with the data points in Fig.~\ref{ivczoom}
would strengthen the evidence for an increase of 
the cooling rate with increasing $N($\ion{H}{1}) in the IV Arch. 

The IVC at positive velocities toward PKS\,2005--489 is only
detected in the metal absorption lines. Therefore, we can only make
a rough estimate of the cooling rate using \ion{P}{2}, assuming
no depletion affects P in this IVC. The cooling rate per nucleon 
inferred from \ion{P}{2} is $-25.28$ dex, the highest cooling rate
in our sample (including the LVCs) for an H column density of $\sim 19.80$ dex
(also inferred from \ion{P}{2}). 

The largest deviation from the mean cooling rate of the LVCs 
is observed in IVC component 2 at $+40$ \km\ toward 
Ton\,S180, where $\log l_c \simeq -24.80$ dex, nearly 8 times
higher than $\langle l_c \rangle$ of the LVCs. The \ion{H}{1} column density
toward Ton\,S180 (component 2) is small, 18.61 dex, 
and for such small column density, there must be an appreciable
amount of \ion{H}{2} along the sightline. 
We do not have information
from the other ions for this component to better characterize 
its properties. \citet{lehner03} find higher cooling rates
per H atom in the LISM at similar 
\ion{H}{1} column densities (see Fig.~\ref{cool}).
The presence of a substantial amount of \ion{H}{2} and 
\ion{C}{2*} in the ionized gas is the most
likely explanation \citep{lehner03}.

Only $3\sigma$ upper limits were derived for IVC-K, IVC-gp, and the IVC
toward NGC\,1068. The high limits toward MRC\,2251--178 (IVC-gp) and
Mrk\,478 (IVC-K) are not really stringent because of a combination of 
low S/N and low \ion{H}{1} column density. The limit toward NGC\,1068 is 
within the observed range of cooling rates.

\subsection{High-Velocity Clouds}\label{hvcdist}
We find a cooling rate of about $-26.99$ dex in Complex C
toward PG\,1259+593 (per H atom or per
nucleon, see Table~\ref{t3}), which is 20 times smaller
than the mean cooling rate of the LVCs. Only
$3 \sigma$ upper limits were derived 
for the other sightlines through the Complex (see Table~\ref{t3}).
Toward Mrk\,817, the upper limit
on the cooling ratio is at least 4 times smaller than the 
Galactic LVC mean value, suggesting that the cooling 
is generally weak in Complex C. The other sightlines do 
not provide stringent limits because the S/N ratio is not high
enough. 

Complex C has a low metallicity (see \S~\ref{hvcgeneral}). 
If C follows a similar abundance
pattern to that of Fe, S, and Si ([C/Si$]\sim -0.1$ dex, Fox et al. 2004), 
it would also be underabundant with respect to the solar abundance by about 
a factor 5--6. Yet,
the cooling rate toward PG\,1259+593 is more than 20 times
smaller than the mean value observed in Galactic Halo gas.
A lower C abundance can only explain in part why the 
cooling rate is so low in Complex C.
Heating and cooling from dust in Complex C cannot
be important because there is no evidence of dust. 
The weak emissivity of \ion{C}{2} in Complex
C with respect to the LVCs and IVCs 
could be a signature of a weaker FUV field. 
Or it could be that the cooling from 
[\ion{C}{2}] emission is a not major 
cooling process because of a mixture of lower metallicity,
warmer gas, and more neutral gas than sampled in the LVCs. 
In the WNM,
\citet{wolfire03} show that for $ 500 \la T \la 8000 $ K
the cooling due to [\ion{O}{1}] at 63 $\micron$ becomes more important, 
and collisional excitation of Ly$\alpha$ (independent of 
the metallicity) dominates the cooling if 
$T \ga 8000$ K. In the pure
WNM, the cooling from [\ion{C}{2}] emission is estimated
to be a factor $\sim$10 lower than in the WIM \citep{wolfire95}.
But while the metallicity is indeed low in Complex C,
there is no evidence that it has warmer or more neutral gas than the LVCs
or IVCs.  The FHWM of the \ion{H}{1} emission 
profile implies a temperature $T \la 8800$ K. 
Complex C is also known to contain ionized gas, as revealed
through H$\alpha$ emission \citep{tufte98,wakker99}. 
\citet{wakker99} derived a temperature in Complex C of 
$T = 7300 \pm 900 \pm^{1500}_{1000}$ K toward Mrk\,290.

We explore another kind of HVC, the Outer Spiral Arm region 
of the Galaxy. We have only a $3\sigma$ upper 
limit for the cooling rate per H atom:  $-25.86$ dex (see Table~\ref{t3}, the
H\,1821+643 sightline). This limit is below 
the mean value observed in Galactic Halo
gas, and in particular for similar or lower \ion{H}{1} column density
observed in the local gas 
\citep[][see Fig.~\ref{cool}]{gry92,lehner03}. We note 
that our $3\sigma$ upper limits are estimated without taking into
account the error on the \ion{H}{1} column density. However, 
toward H\,1821+643 in particular, the $1\sigma$ error is
small, only $\pm 0.06$ dex. Hence, the derived upper limits
are lower than the cooling rate  for lower-velocity components in the Galaxy.

While Complex C and the Outer Arm (OA) are unrelated HVCs, they both
contain little or no dust, have ionized gas
kinematically related to the neutral gas, and possibly
trace exclusively warm gas ($T \sim 7000-8000$ K). 
They are also distant. Complex C is believed to be extragalactic and at 
least 6 kpc away from the Milky Way disk. 
The OA HVC could be at a galactocentric distance of $\sim$24 kpc.  
This may imply that the FUV 
radiation field in these HVCs is substantially weaker, and hence other
sources of heating from EUV and X-rays may be important. 
A possible signature of EUV and (soft) X-ray radiation is \ion{O}{6} absorption detected 
toward these sightlines at similar velocities as
the low ionization species. In particular, \citet{fox04} showed
that the high ions in Complex C toward PG\,1259+593
are probably produced in an interface between cool/warm gas and a surrounding 
hot medium. The interface is a possible source of EUV radiation and
the surrounding hot gas is a source of X-ray radiation. 

Toward Mrk\,205, the very high-velocity cloud WW84 also gives
a $3\sigma$ upper limit 1.5 times smaller than the mean value
observed in our Galaxy. 

\subsection{Comparison with Other (More Local) UV Observations}
\citet{pottasch79} measured the \ion{C}{2} cooling rates per nucleon
for the interstellar gas outside dense \ion{H}{2} regions toward
9 early-type stars. Their study was based on {\em Copernicus} and
{\em IUE} observations. They found that the cooling rates 
did not appear to vary substantially from one cloud 
to  another, although they did not produce any error estimates. 
They found a mean cooling rate around $-25$ dex (per nucleon). Their mean 
is indicated by horizontal dashed line in Fig.~\ref{cool}. It is
substantially higher than most of our measurements and the results of other studies
at similar \ion{H}{1} column densities. Their cooling rates may be 
biased toward high values since the FUV radiation fields
appear to be greater toward at least some of their sightlines 
than the interstellar average \citep{wolfire95}.

\citet{gry92} extended the \citet{pottasch79} study by gathering {\em Copernicus}
observations to measure \ion{C}{2*}
column densities  toward 20 stars that are situated at distances of a few hundred pc
to 1.4 kpc.  They found a dispersion larger than the errors in the cooling
rates from direction to direction, with a mean value of $-25.46 \pm 0.41$ dex
(per nucleon), smaller by 0.46 dex than the value of  \citet{pottasch79}, but higher
by 0.24 dex than along the Galactic halo sightlines presented in this study (see the summary
Table~\ref{t5}). The magnitude of the scatter of the cooling rates is larger 
than the one observed in our survey, but note that several 
of their measurements have $1\sigma$ error larger than $>\pm 0.7$ dex  
(see Fig.~\ref{cool}). Because 
their background FUV sources are early-type stars, \citet{gry92}
expected that some of the \ion{C}{2*} cooling takes place in the \ion{H}{2} 
regions which surround the stars. This would explain the higher mean 
cooling rate in their survey. Cold gas is also more
likely to be present along their sightlines. 

\citet{lehner03} derived the cooling rates in the LISM
toward 31 white dwarf stars located less than 200 pc from the sun.
If we exclude two of their sightlines for which substantial
ionization corrections are needed, their mean cooling rate per H atom is
$l_c = -25.63 \pm\,^{0.18}_{0.33}$ dex. Their mean and dispersion
of the cooling rate in the LISM are very similar to what we find for the 
Galactic Halo interstellar clouds (see Table~\ref{t5} and Fig.~\ref{cool}), suggesting 
a similarity of the physical properties of the LISM  and the Galactic 
Halo gas.

\subsection{Comparison with Galactic Halo IR Observations}
Another method to obtain the \ion{C}{2} cooling rate per H atom is from direct measurements
of the [\ion{C}{2}] 157.7 $\micron$ line via IR observations.
\citet{bock93} measured the [\ion{C}{2}] emission line toward three directions between  $l = 135\degr-150\degr$ 
and $b = 33\degr-50\degr$. The emission of [\ion{C}{2}] is observed in all directions, it correlates well with $N($\ion{H}{1}). 
They found a cooling rate that spans values from 
$l_c = -25.92$ to $-25.49$ dex (for $N($\ion{H}{1}$)>  10^{20}$ cm$^{-2}$),
with a best-fit value of$-25.58 \pm 0.10$ dex, excluding the cases with low line-to-continuum ratios
and the lines of sight with CO emission. \citet{matsuhara97}, using the same data but concentrating on 
high Galactic latitude molecular clouds, found  $l_c = -25.80 \pm 0.11$ dex
for $N($\ion{H}{1}$)<  2 \times 10^{20}$ cm$^{-2}$.
At high galactic latitude and on nearly the full Galactic sky
with two instruments (FIRAS and DIRBE) onboard of the {\em Cosmic Background
Explorer} ({\em COBE}) \citet{bennett94} found a best-fit value
of $l_c = -25.57 \pm 0.03 $ dex, while \citet{caux97} found $l_c = -25.48 \pm 0.04 $ dex
with the {\em Infrared Space Observatory} ({\em ISO}). Note that the
errors reported in the FIR studies are the errors on the fit between 
the [\ion{C}{2}] 157.7 $\micron$ intensity and the \ion{H}{1} column
density, and not the dispersion of the measurements. 

Within the IR measurements there is some scatter between 
the different best-fit values of $l_c$. But 
the range of cooling rate values and the best-fit values reported
in the IR studies at high galactic latitudes are in good agreement with the mean 
cooling rate and dispersion derived in our survey for the Galactic halo gas (LVCs), which
is important in view of the fact that these two methods are completely
different. \citet{bennett94}
argued that this emission arises almost entirely from cold regions. 
\citet{makiuti02} using the Far-Infrared
Line Mapper (FILM) aboard of the {\em Infrared Telescope in Space} ({\em IRTS})
recently argued that [\ion{C}{2}] emission must mostly come from the WIM at 
high Galactic latitudes. 
Our analysis also shows that at $\mid\! b\! \mid \ga 30 \degr$ 
a large fraction of [\ion{C}{2}] comes from the WIM (see \S~\ref{lvcdiscuss}).
We also note that \citet{heiles94} shows that the ionized medium could explain 
the bulk of [\ion{C}{2}] in the inner Galaxy.

\section{Other Implications of the \ion{C}{2*} Absorption}\label{otherimplication}

\subsection{The Origin of C\,{\sc ii*} at High Galactic Latitude}\label{lvcdiscuss}

The column density of \ion{C}{2*} in the WIM can be estimated in the following way 
(R. J. Reynolds 2004, private communication): In \S~\ref{elecdens} we discuss that
in the WIM and WNM conditions,  Eq.~\ref{elfulleqt} can be simplified, and so can 
be written in the WIM as:
\begin{equation}\label{colsimp}
n({\rm C^{+*}}) A_{21} = n({\rm C^+}) n_e \gamma_{12}(e)	
\end{equation}
We can integrate Eq.~\ref{colsimp} over the line of sight as follows:
$$
\int \alpha{({\rm H \alpha})}  A_{21}  n({\rm C^{+*}})  {\rm d}s
= 
10^6 \int \frac{n({\rm C^+})}{n({\rm C})}\frac{n({\rm C})}{n({\rm H})}
\frac{n({\rm H})}{n({\rm H^+})}\gamma_{12}(e)   \frac{\alpha{({\rm H \alpha})}}{10^6} n_e n({\rm H^+}) {\rm d}s  \,,
$$
or, assuming that the fractions of \ion{C}{2}, C, and \ion{H}{2} are constant (see below),
$$ 
\alpha{({\rm H \alpha})}  A_{21} N_{\rm WIM}({\rm C^{+*}})
= 
10^6 \frac{n({\rm C^+})}{n({\rm C})}\frac{n({\rm C})}{n({\rm H})}
\frac{n({\rm H})}{n({\rm H^+})}\gamma_{12}(e) I[{\rm H \alpha}] \,,
$$
where
$$
I[{\rm H \alpha}] = \int \frac{\alpha{({\rm H \alpha})}}{10^6} n_e n({\rm H^+}) {\rm d}s
$$
is the velocity-integrated surface brightness of 
diffuse H$\alpha$ emission in Rayleigh ($1 {\rm R} \tbond 10^6/4\pi$  
photons cm$^{-2}$\,s$^{-1}$\,sr$^{-1}$). $\alpha{({\rm H \alpha})} = 1.17\times 10^{13}$
cm$^3$\,s$^{-1}$ \citep{martin88,osterbrock89} is the recombination coefficient 
of H$\alpha$ at $T = 10^4$ K, and all the other symbols are defined in 
\S~\ref{elecdens}. We assume the temperature in the WIM $T = 10^4$ K
based on \citet{reynolds93} study. Assuming that \ion{C}{2} is the dominant ion in the WIM
(Sembach et al. 2000 modeled the WIM and showed that \ion{C}{3} is negligible with respect to \ion{C}{2}, and
see also Reynolds 1992), $n({\rm C^+})/n({\rm C}) \simeq 1$. Assuming that the gas-phase abundance
of C is the same in the diffuse CNM, WNM and the WIM, $n({\rm C})/n({\rm H}) \simeq 1.42 \times 10^{-4}$
\citep{sofia97}. Finally, in the WIM, $n({\rm H})/n({\rm H^+}) \simeq 1.1$ \citep{reynolds89}. Then
the column density of \ion{C}{2*} in the WIM is simply given by,
\begin{equation}\label{eqtwim}
N_{\rm WIM}({\rm C^{+*}}) = 7.23 \times 10^{13} I[{\rm H} \alpha] \,\, {\rm cm}^{-2}\,.
\end{equation}
The $I[{\rm H \alpha}]$ intensity was measured by the Wisconsin H$\alpha$ 
mapper (WHAM) \citep{haffner03} over a $1\degr$ diameter field of view
that contains  the background object, but not centered on the object.  In
Table~\ref{t5a}, we list the sightlines for which there is a $I[{\rm H \alpha}]$
measurement, where $\Delta \theta$ is the distance in degrees to the nearest H$\alpha$  
survey gridpoint. For HD\,93531, we list the results from the WHAM survey as well as 
the direct pointing of WHAM on that object, in which the LVC and IVC components were separated 
\citep{hausen02}. In this table, $N_{\rm obs}($\ion{C}{2*}) is the observed \ion{C}{2*}
column density that includes the LVC and the IVC components, because $I[{\rm H \alpha}]$
is integrated over $\sim\pm 100$ \km\ in the WHAM survey. We did not include 
the measures for which $N_{\rm obs}($\ion{C}{2*}) is a lower limit or uncertain. 
We also did not include the value of $I[{\rm H \alpha}] = 18.6$ R toward Mrk\,1095
because such a high value is certainly contaminated by a dense \ion{H}{2} region not
associated with the diffuse WIM and WNM (this line of sight lies the closest 
to the galactic plane, at $b\simeq -21\degr$). In Table~\ref{t5a} we tabulate the estimated
values of $N_{\rm WIM}($\ion{C}{2*}) and $N_{\rm WIM}($\ion{C}{2*}$)/N_{\rm obs}($\ion{C}{2*}).
We show in Fig.~\ref{ciiwim} $N_{\rm WIM}($\ion{C}{2*}) against $N_{\rm obs}($\ion{C}{2*}). 

The fraction $N_{\rm WIM}($\ion{C}{2*}$)/N_{\rm obs}($\ion{C}{2*}) varies significantly 
from sightline to sightline between 0.1 to 1. The mean and  median values
of $N_{\rm WIM}($\ion{C}{2*}$)/N_{\rm obs}($\ion{C}{2*}) both are 0.5,
with a dispersion of 0.2. This calculation contains
uncertainties. The gas-phase abundance of C in the WIM 
may be different than in the diffuse CNM. If the abundance of C in the WIM
is solar, this would increase $N_{\rm WIM}($\ion{C}{2*}) by a factor
1.73, and in that case the mean of 
$N_{\rm WIM}($\ion{C}{2*}$)/N_{\rm obs}($\ion{C}{2*}) would be
0.8.  If the temperature of the WIM is 
8000 K instead of $10^4$ K, $N_{\rm WIM}($\ion{C}{2*}) would
increase by about 10\%. However, the fractions $n({\rm C^+})/n({\rm C}) \simeq 1$
and $n({\rm H})/n({\rm H^+}) \simeq 1.1$ should not change by much more 
than 5--10\% in the WIM \citep{reynolds89,sembach00}. 
Hence, the uncertainties of the 
different factors used to estimate $N_{\rm WIM}($\ion{C}{2*}) show that
the fraction of \ion{C}{2*} in the WIM is at least 0.5. 

Another uncertainty is that the gas
sampled by WHAM and the FUV measures may not be the same
because of the 1\degr\ field of view of WHAM compared
to the very small angles subtended by the AGNs or stars.
The effect of this uncertainty can not be easily quantified. Toward
HD\,93521, $N_{\rm WIM}($\ion{C}{2*}) does not change much
between a direct pointing on that object and the 0\fdg41 distant
pointing, at least for 
the LVC component. On the other hand, toward vZ\,1128 we know that
the fraction of ionized hydrogen is 46\% and the ionized and
neutral phases are kinematically related \citep{howk04}, 
so it is surprising that only 13\% of $N_{\rm obs}($\ion{C}{2*}) comes
from the WIM based on the H$\alpha$ estimates. We note that for another pointing 
0\fdg6 away from vZ\,1128, $I[{\rm H \alpha}]$
is  2.8 times the value of $I[{\rm H \alpha}]$ given in Table~\ref{t5a}.
In contrast, toward both NGC\,5904-ZNG1 and NGC\,6205-ZNG1, where 20\% and
40\% of hydrogen is ionized (see \S~\ref{lvcintro}), 
at least 50\% of $N_{\rm obs}($\ion{C}{2*}) is from the WIM. 
$I[{\rm H \alpha}]$ should, however, not be systematically high 
or low, so that if the individual sightline may suffer from some 
uncertainty because of the 1\degr\ field of view of WHAM, 
the average value of  $N_{\rm WIM}($\ion{C}{2*}$)/N_{\rm obs}($\ion{C}{2*}) should 
not be affected by the irregular distribution of the gas.

\subsection{The Integrated Galactic \ion{C}{2} Radiative Cooling Rate}\label{intcool}
Since we know the \ion{C}{2} radiative cooling rate per H atom and per nucleon toward many
sightlines in the Milky Way, we can  make a rough estimate of the 
total \ion{C}{2} cooling rate for the diffuse gas in the entire Galaxy.
If we assume an exponential disk, the total number of warm H atoms 
in the Galaxy can be written:  
\begin{equation}
N({\rm atoms})  =  \int_0^{2\pi}\int_{-\infty}^{+\infty} \int_0^{R_{\rm max}}  
		n(0)\, e^{(-z/h)}\, e^{(-R/H)}\, R \,dR\, dz \, d\phi  \,,
\end{equation}
where $H $ and $h$ are the scale lengths and scale heights, respectively, $R_{\rm max} \simeq 25$
kpc \citep{kulkarni82,diplas91}, and $n(0)(=n(R_\odot)e^{(R_\odot/H)})$ is the mid-plane density at the galactic center. Assuming that 
$h$ is independent of $R$, the total number of warm H atoms in the 
Milky Way is then
\begin{eqnarray}
N({\rm atoms}) & = &2 
	\pi e^{(R_\odot/H)} H^2 N_\perp  \int_0^{R_{\rm max}}  e^{(-R/H)} (R/H){\rm d}(R/H) \,  \nonumber \\
	  & \simeq & 2 \pi e^{(R_\odot/H)} H^2 N_\perp \,,
\end{eqnarray}
where $H \simeq 3$ kpc \citep{diplas91}  and $R_\odot = 8.5$ kpc. We estimated 
the total average perpendicular column density of neutral hydrogen for the WNM
measured at the position of the Sun, $R_\odot$, from the Leiden/Dwingeloo \ion{H}{1} 
survey \citep{hartmann97},
$N_\perp ($\ion{H}{1}$) = 2 \times (2\times10^{20})$ cm$^{-2}$.
Taking into account that the total average perpendicular column density for  the WIM 
at $R_\odot$ is $ N_\perp ($\ion{H}{2}$) = 2 \times (7.1\times10^{19})$ cm$^{-2}$ \citep{reynolds93},
the total perpendicular
H column density at $R_\odot$ is $5.4 \times 10^{20}$ cm$^{-2}$. So, with the mean
cooling rate of the LVCs, the total luminosity associated with the \ion{C}{2} cooling 
in the Galaxy from the WNM and WIM (where we assume that the local
conditions apply to the entire Galaxy) is $L \simeq 9.9 \times 10^{40}$ erg\,s$^{-1}$ or 
$L \simeq 2.6 \times 10^7$ L$_\odot$. \citet{wright91} found a higher 
value with {\em COBE}, $L \simeq 5.0 \times 10^7$ L$_\odot$.  \citet{shibai91}
found a similar value ($L \simeq 2.7 \times 10^7$ L$_\odot$), using a balloon
experiment along the galactic plane for $21\degr \le l\le 51 \degr$.

\subsection{Electron Density}\label{elmeas}
Using  Eq.~\ref{elredeqt} (at $T=6000 $ K, 
$n_e \simeq 15.29 N({\rm C}^{+*})/N({\rm C}^{+})$ cm$^{-3}$)
our data also allow estimates of the electron density
in the gas. 
The \ion{C}{2} $\lambda$$\lambda$1036, 1334 absorption lines  are, however, 
extremely strong, and the \ion{C}{2} column density 
can not be estimated reliably. C is lightly 
depleted into dust grains ($-0.2$ dex, Cardelli et al. 1996; Sofia et al. 1997; Jenkins 2003).  
Therefore we can use \ion{S}{2} as a proxy for \ion{C}{2}.
Only a small percentage, if any, of S is
incorporated into dust grains, but C is lightly depleted by 
$-0.2$ dex, which needs to be taken into account. 
The first ionization stage 
of all these species is also the dominant one in warm gas.
So, we can assume 
$N($\ion{C}{2*}$)/N($\ion{C}{2}$) \approx 10^{0.2} 
N($\ion{C}{2*}$)/N($\ion{S}{2}$) \times ({\rm S}/{\rm C})_\odot$, 
where the solar ratio $\log({\rm S}/{\rm C})_\odot = -1.06$ dex \citep{grevesse98,allende02}. 
If \ion{C}{2} is mostly in \ion{H}{1} regions,
we can also use our \ion{H}{1} measurement, the 
\citet{allende02} C solar abundance ($\log({\rm C}/{\rm H})_\odot = -3.61$) 
and a depletion of $-0.2$ dex of C to estimate $n_e$.  
Note that we did not make any correction for depletion effects in the HVCs. For
Complex C and WW84 we have corrected for the metallicity when \ion{H}{1}
is used as a proxy for \ion{C}{2}.  

The electron densities  are summarized in Table~\ref{t3} and displayed 
for $n_e($\ion{H}{1}) in Fig.~\ref{cool}  against $N($\ion{H}{1}), where a temperature
of the gas of 6000 K was assumed. 
Because $n_e \propto l_c$, the main discussion in \S~\ref{coolmeas} 
about variations in the cooling rates and the origin of \ion{C}{2*}  directly 
applies to $n_e$. Here, we summarize the mean value of 
$n_e$ and dispersion for the different type of clouds investigated.

{\em The LVCs:} We find a mean value of the electron density
(using \ion{H}{1} or \ion{S}{2}) of 
$\langle  n_e \rangle = 0.08 \pm 0.04 $ cm$^{-3}$ when $T=6000$ K, 
and a range of values, $ 0.02 \la n_e \la 0.17$ cm$^{-3}$. 
The only value departing from this range is $n_e \simeq 0.006$ cm$^{-3}$ 
toward Mrk\,1095, confirming that this sightline is predominantly neutral. 
In the LISM and in the high galactic latitude diffuse clouds from other FUV measurements, 
the range of measured electron densities was found to be 
$ 0.01 \la n_e \la 0.15$ cm$^{-3}$ for temperatures of about
6000--7000 K, with an average value of about 0.07 cm$^{-3}$ 
\citep[e.g.,][]{savage93,spitzer93,spitzer95,savage96,gry01,lehner03}. 
The mean, dispersion, and range are thus similar to those found in previous FUV studies, 
but now with a better statistic at high galactic latitudes. 
The electron density has also been inferred from a comparison of the 
H$\alpha$ emission with the dispersion measure
for pulsars in globular clusters. Using this method, \citet{reynolds91} 
found also an average electron density of 0.08 cm$^{-3}$ toward four  
high Galactic latitude globular clusters (this work includes two directions 
in our sample, NGC\,5904 and NGC\,6205).

{\em The IVCs:} In the IV Arch, we have 
$\langle  n_e \rangle \simeq 0.03 \pm 0.01 $ cm$^{-3}$ when $T=6000$ K, 
and a range of values of $ 0.01 \la n_e \la 0.05$ cm$^{-3}$. 
The other IVCs have similar electron density (see Table~\ref{t3}), 
except toward Ton\,S180 ($n_e \simeq 0.54$ cm$^{-3}$) but for which
a substantial ionization correction is necessary because $n_e$ was
derived using the \ion{H}{1} column density. For the IVC toward 
HD\,93521, \citet{spitzer93} also reported a similar electron density. 

{\em The HVCs:} Only one tentative measurement is derived, 
toward PG\,1259+593, giving $n_e \sim 0.01$ cm$^{-3}$.  
The $3\sigma$ upper limits toward the other targets 
are in the range observed for the lower velocity clouds. 

\subsection{Damped Ly\,$\alpha$ Systems}\label{dlacomp}
In Fig.~\ref{cool} we compare our results 
to the recent survey of \ion{C}{2} cooling
rates in the damped Ly\,$\alpha$ systems (DLAs) reported 
by \citet{wolfe03a}.  
The mean cooling rate and dispersion around the mean are
summarized in Table~\ref{t5}. A comparison of the means reveals 
that $\log \langle l_c \rangle$ in the DLAs
is $\sim$1 dex less than value of $\log \langle l_c \rangle$
of the LVCs in the Galactic Halo. For
$\log N($\ion{H}{1}$)\ga 20.5$ dex, the difference with 
the Galactic Halo sightlines is smaller. The dispersions 
of the cooling rates are similar in both samples.  We discussed
that the variation in the Milky Way cooling rate has certainly several origins,
including changes in the ionization fraction, in the dust fraction, 
and in the physical conditions. In the DLAs, 
\citet{wolfe03a}  proposed that
the variation is due to different star-formation rates (SFRs).
In order to derive the SFR in the DLAs, the models of \citet{wolfe03a} 
assume that  the reservoir of \ion{C}{2*} in the DLAs gas 
comes from the CNM. Recently, \citet{howk04a} directly show that
cold gas is likely the dominant contributor to \ion{C}{2*} for one DLA of their sample. However,
while \citet{vladilo01} (using the ratio of \ion{Al}{3}/\ion{Al}{2})
showed that the ionization correction for elemental abundance analysis may be
unimportant, the ratio of \ion{Al}{3}/\ion{Al}{2} in most DLAs
implies nonetheless a significant fraction of ionized gas in the 
DLAs, similar to that observed in the Milky Way. The presence
of a WIM component in the DLAs could be an important contribution
to the \ion{C}{2*} in DLAs.

We also note the similarity of the measured cooling rates in 
Complex C and the DLAs. Complex C 
and DLAs both contain low-metallicity gas,
but in Complex C there is no evidence for stars or dust. 
Also \citet{wolfe03a} discussed that the main gas
phase observed in the DLAs may be cold and neutral, while 
it is warm neutral and ionized in Complex C. Yet, the similar \ion{C}{2} cooling 
rates in Complex C and DLAs suggest that some 
of the DLAs could be intergalactic clouds near galaxies like 
the gas in Complex C, 
rather than clouds in which stars are currently forming.

\section{Summary}\label{summary}
We present a survey of the [\ion{C}{2}] $\lambda$157.7 $\micron$  
radiative cooling rates from the \ion{C}{2*} $\lambda$$\lambda$1037, 1335 
absorption lines at galactic latitudes $\mid\!  b\!  \mid  \ga 30 \degr$
using  {\em FUSE}\/ and STIS observations. Our survey allows 
us to derive the \ion{C}{2} cooling rates in the low-, intermediate-,
and high-velocity clouds (LVCs, IVCs, HVCs). The main results 
are summarized in Tables~\ref{t3} and \ref{t5}, and in Fig.~\ref{cool}
and are discussed in \S~\ref{coolmeas}.
Our main conclusions are summarized as follows: 

\begin{enumerate}
\item For the LVCs, the logarithm of mean cooling rate in erg\,s$^{-1}$ per H atom is
$-25.70^{+0.19}_{-0.36}$ dex ($1\sigma$ dispersion). With a smaller
sample and a sample bias toward \ion{H}{1} column densities larger than
$10^{20}$ cm$^{-2}$, the cooling rate per nucleon is similar.
\item Our sightlines probe mostly warm clouds based on measures
of the depletion of the Fe and P into dust.
We are able to show that a substantial fraction of hydrogen is
ionized (20--50\%) toward a few of our sightlines, and argue
that ionization is certainly important toward most of them. 
We find that a fraction  of at least 0.5 
of the observed \ion{C}{2*} is produced in the WIM, using H$\alpha$ measurements in the 
direction of each object.
\item The observed dispersion in the cooling rates is larger than the individual 
measurement errors. The dispersion certainly arises from
changes from sightline to sightline in the ionization fraction, dust-to-gas fraction, 
and physical conditions.
\item We derive a total Galactic \ion{C}{2} luminosity for gas in the WNM and WIM 
of $L \sim 2.6 \times 10^7$ L$_\odot$.
\item The mean and dispersion of the 
cooling rates of the LVCs observed at high Galactic latitudes 
in our survey are very similar to the cooling rates observed in the local ISM
within 200 pc. The mean cooling rate is, however, a factor $\sim$2 lower than 
the FIR measurements at high latitudes, but compatible within our $1\sigma$ 
error. 
\item The main IVC probed in our survey is the IV Arch located at 
$z\sim 1$ kpc. The logarithm of the mean cooling rate per nucleon is $-25.98$ dex with a
dispersion of about $\pm\,^{0.14}_{0.20}$ dex. This value 
is 2 times smaller than the mean cooling of the LVCs, implying 
the cooling and hence the heating decrease at higher $z$. 
\item Two IVCs toward PKS\,2005--489 and  Ton\,S180 
with positive velocity for which the distance
is unknown possess high cooling rates. 
\item The observations of \ion{C}{2*} yield estimates of $n_e$ in the absorbing gas.
Assuming $T = 6000$ K, for the LVCs, $\langle  n_e \rangle = 0.08 \pm 0.04 $ cm$^{-3}$ and 
for the IV Arch, $\langle  n_e \rangle = 0.03 \pm 0.01 $ cm$^{-3}$
($1\sigma$ dispersion).
\item At large $z$ ($> 6$ kpc), the \ion{C}{2} cooling rate of the gas in Complex C 
is 20 times smaller than the average for the LVCs, implying that
the heating and cooling processes are far weaker at large distance from the 
Milky Way disk.
\item  The \ion{C}{2} cooling rate found for high redshift DLAs by \citet{wolfe03a}
is $\sim$1 dex smaller than for the Galactic halo LVCs, but comparable to the 
cooling rate for the gas in Complex C. 
\item Many of the DLAs have values of \ion{Al}{3}/\ion{Al}{2} similar to that 
found in the Milky Way, suggesting roughly similar amounts of ionized gas to
neutral gas. Therefore, we would expect ionized gas in many of DLAs to provide
an important contribution to \ion{C}{2} cooling. 
\end{enumerate}

\acknowledgments
We thank Ron Reynolds for his idea on how to estimate the fraction 
of \ion{C}{2*} in  the WIM. We also thank Chris Howk and Art Wolfe
for useful discussions. We are grateful to Marylin 
Meade for reprocessing the {\em FUSE}\/ data.
We thank the anonymous referee for providing 
a number of important comments and suggestions about the
original manuscript. 
This work is based on observations made with the NASA/ESA 
{\em Hubble Space Telescope}, obtained from the Data
Archive at the Space Telescope Science Institute, 
which is operated by the Association of Universities for
Research in Astronomy, Inc., under NASA contract NAS 5-26555. 
Support for MAST for non-HST data is 
provided by the NASA Office of Space Science via grant NAG5-7584 and 
by other grants and contracts.  
Based on observations made with the NASA-CNES-CSA {\em Far Ultraviolet 
Spectroscopic Explorer}. {\em FUSE}\/ is operated for NASA by the 
Johns Hopkins University under NASA contract NAS5-32985.
This research has made use of the NASA
Astrophysics Data System Abstract Service and the SIMBAD database,
operated at CDS, Strasbourg, France.
This work has been supported
by NASA through grants NNG-04GC70G (BDS),  NAG5-9179 and
NAG5-13687 (BPW).

\begin{deluxetable}{lccccccccll}        
\tabcolsep=2pt
\tablecolumns{11}
\tablewidth{0pt} 
\tabletypesize{\tiny}
\tablecaption{Basic Data for Extragalactic Sightlines \label{t1}}
\tablehead{\colhead{Name} &\colhead{R.A.} &\colhead{Dec.} &\colhead{$l$} &\colhead{$b$}      &\colhead{redshift}&  \colhead{Type}  &\colhead{$B$}& \colhead{$E(B-V)$}&\colhead{Exp. Time}&\colhead{Instrument} \\
\colhead{}&\colhead{(h\ m\ s)}&\colhead{($\circ$\  '\  ")}&\colhead{($\circ$)}&\colhead{($\circ$)}&\colhead{}&\colhead{}&\colhead{(mag)}&\colhead{(mag)} &\colhead{(ks)}  &\colhead{} \\
\colhead{(1)}& \colhead{(2)} & \colhead{(3)}& \colhead{(4)}& \colhead{(5)}&\colhead{(6)}&\colhead{(7)}&\colhead{(8)}&\colhead{(9)} &\colhead{(10)} &\colhead{(11)}}
\startdata 
3C\,273.0          &12 29 06.7&+02 03 08	&289.95& +64.36	&0.1583&QSO   			&13.05&0.021  & 43(F), 19(S)   	&{\fuse}, STIS, GB	  \\
3C\,351.0          &17 04 41.5&+60 44 28	& 90.08& +36.38	&0.3719&QSO   			&15.41&0.023  & 77(S)		&STIS, Eff		  \\
H\,1821+643        &18 21 57.2&+64 20 36	& 94.00& +27.42	&0.2844&QSO   			&14.23&0.043  & 51(S)		&STIS, Eff		  \\
HE\,0226--4110     &02 28 15.2&$-$40 57 16&253.94&$-$65.77&0.4949&QSO   			&14.23&0.043  & 112(F), 44(S)   &{\fuse}, STIS, VE		  \\
HE\,1228+0131      &12 30 50.0&+01 15 21  &291.26& +63.66 &0.1170&QSO   			&14.44&0.019  & 27(S)   	&STIS, LDS		  \\
MRC\,2251--178     &22 54 05.9&$-$17 34 55& 46.20&$-$61.33&0.0660&QSO   			&14.99&0.039  & 45(F)   	&{\fuse}, GB  \\
Mrk\,54            &12 56 55.7&+32 26 51  &110.64& +84.55 &0.0447&Gal:Sc			&15.26&0.015  & 27(F)   	&{\fuse}, LDS  \\
Mrk\,59            &12 59 00.3&+34 50 44	&111.54& +82.12	&  1089&HII(NGC4861)		&13.65&0.011  & 10(F)   	&{\fuse}, LDS  \\
\enddata
\tablecomments{
Column 6: Redshift if less than unity, or in units of velocity (\km) if greater than unity. 
Column 10:  "F" stands for {\fuse}, "S" stands for STIS. For {\fuse},
the exposure time is based on the LiF\,1A segment. Other segments may have smaller exposure time in some cases. 
Column 11: Instrument used. 
Eff: Effelsberg telescope ($9\arcmin$ beam),
GB: Green Bank telescope ($21\arcmin$ beam);
LDS: Leiden-Dwingeloo Survey ($35\arcmin$ beam);
VE: Villa Elisa telescope ($34\arcmin$ beam).  
}
\end{deluxetable}

\begin{deluxetable}{lcccccclccll}        
\tabcolsep=2pt
\tablecolumns{12}
\tablewidth{0pt} 
\tabletypesize{\tiny}
\tablecaption{Basic Data for Galactic Sightlines \label{t1a}}
\tablehead{\colhead{Name} &\colhead{R.A.} &\colhead{Dec.} &\colhead{$l$} &\colhead{$b$}      &\colhead{$d$} &\colhead{$z$}&  \colhead{Type}  &\colhead{$B$}& \colhead{$E(B-V)$}&\colhead{Exp. Time}&\colhead{Instrument} \\
\colhead{}&\colhead{(h\ m\ s)}&\colhead{($\circ$\  '\  ")}&\colhead{($\circ$)}&\colhead{($\circ$)}&\colhead{(kpc)}&\colhead{(kpc)}&\colhead{(mag)}&\colhead{}&\colhead{(mag)}  &\colhead{(ks)} &\colhead{}\\
\colhead{(1)}& \colhead{(2)} & \colhead{(3)}& \colhead{(4)}& \colhead{(5)}&\colhead{(6)}&\colhead{(7)}&\colhead{(8)}&\colhead{(9)} &\colhead{(10)} &\colhead{(11)} &\colhead{(12)}}
\startdata 
HD\,18100          &02 53 40.8&$-$26 09 20&217.93 &$-$62.73 &2.0  &$-2.2$  & B1\,V	  &8.46 &0.020   & 1.9(F)  	    &{\fuse}, GHRS, LDS \\
HD\,233622         &09 21 33.6&+50 05 56  &168.17 &+44.23   &4.4  &+3.1    & B2\,V	  &10.01&0.030   & 4.7(F), 1.4(S)   &{\fuse}, STIS, Eff   \\
HD\,93521          &10 48 23.5&+37 34 13  &183.14 & +62.15  &1.9  &+1.7    & O9.5\,Vp& 7.10&0.040   & 7.3(F)  		    &{\fuse}, GHRS, JB	  \\
HD\,97991          &11 16 11.7&$-$03 28 19&262.34 &+51.73   &1.0  &+0.8    & B1\,V	  &7.13 &0.010   & 0.059(F)  	    &{\fuse}, LDS		\\
NGC\,5904-ZNG1	 &15 18 34.1&+02 04 59	& 3.87  & +46.80  &7.0  &+5.1    & PAGB/GC &14.20&0.030   & 3.7(F)  		    &{\fuse}, LDS     \\
NGC\,6205-ZNG1	 &16 41 34.3&+36 26 08	& 58.97 & +40.93  &6.8  &+4.5    & PAGB/GC &14.90&0.017	 & 15.9(F)		    &{\fuse}, LDS     \\
PG\,1051+501  	 &10 54 18.5&+49 50 01  &159.61 &+58.12   &2.6  & +2.2   & sdB     &14.59& 0.021  & 3.9(F)  		    &{\fuse}, LDS	\\
\enddata
\tablecomments{
Column 6: Distance of the star.
Column 7: $z$-height of the star.
Column 11: Exposure times in ks, "F" stands for {\fuse}, "S" stands for STIS. For {\fuse},
the exposure time is based on the LiF\,1A segment, except for HD\,93521 which is from 
SiC\,1B. Other segments may have smaller exposure time in some cases. 
Column 12: Instrument used. For HD\,18100 and HD\,97991, the H\,{\sc i} Ly\,$\alpha$ column densities 
based on {\em IUE} spectra were used in the calculations (Results from Diplas \& Savage 1994).
For HD\,18100 and HD\,93521, GHRS observations are also used (results from Savage \& Sembach 1996b, Spitzer \& Fitzpatrick 1993, respectively).
Eff: Effelsberg telescope ($9\arcmin$ beam),
GB: Green Bank telescope ($21\arcmin$ beam);
JB: Jodrell Bank telescope ($12\arcmin$  beam);
LDS: Leiden-Dwingeloo Survey ($35\arcmin$ beam);
}
\end{deluxetable}

\begin{deluxetable}{lcccccccccc}
\tabcolsep=1pt
\tablecolumns{11}
\tablewidth{0pt} 
\tabletypesize{\tiny}
\tablecaption{Column densities \label{t2}} 
\tablehead{\colhead{Target}    &   \colhead{$\log N($Fe\,{\sc ii})} &   \colhead{$b($Fe\,{\sc ii})}&   \colhead{$\log N_a($C\,{\sc ii*})}&   \colhead{$\log N($C\,{\sc ii*})} &   \colhead{S/N}&   \colhead{$\log N($S\,{\sc ii})} &   \colhead{$b($S\,{\sc ii})} &   \colhead{$\log N($P\,{\sc ii})} &   \colhead{$\log N($H\,{\sc i})} & \colhead{Components} \\ 
\colhead{}  &\colhead{(dex)} & \colhead{(\km)}&\colhead{(dex)} &\colhead{(dex)}  &\colhead{} &\colhead{(dex)}& \colhead{(\km)}&\colhead{(dex)}&\colhead{(dex)} & \colhead{}\\
\colhead{(1)}& \colhead{(2)} & \colhead{(3)}& \colhead{(4)}& \colhead{(5)}&\colhead{(6)}&\colhead{(7)}&\colhead{(8)}&\colhead{(9)} &\colhead{(10)}&\colhead{(11)}}
\startdata
3C\,273.0		&  $ 14.87 \pm 0.02		 $ &  $  16.5 \pm 0.5		   $	 &  $  13.80 \pm 0.02	$    &  $  13.84 \pm 0.02	      $    &31(S)    &  $  15.25 \pm 0.03   $	       &  $  (14.7\pm\,^{2.2}_{1.7})^b     $	   &  $  13.46 \pm 0.07    $	   & $  20.15 \pm  0.10  $   &   1+2+3 	       \\
3C\,273.0		&  $ 14.46 \pm 0.02		 $ &  $  10.2 \pm  0.5  	   $	 &  $  13.65 \pm 0.02	$    &  $  13.68 \pm 0.03	      $    &31(S)    &  $  14.83 \pm 0.04   $	       &  $  (7.1\pm\,^{2.5}_{1.3})^b	   $	   &  $  13.09 \pm 0.10    $	   & $  19.42 \pm  0.10  $   &   4			     \\
3C\,351.0		&  \nodata			   &  \nodata				 &  $  > 14.14  	$    &  $ 14.28\pm\,^{0.66}_{0.17}\,^a$    &14(S)    &  $  15.62\pm\,^{0.76}_{0.24}$   &  $  25.6\pm\,^{8.4}_{7.9}	   $	   &  \nodata			   & $  20.27 \pm  0.06  $   &   2+3			\\
3C\,351.0		&  \nodata			   &  \nodata				 &  \nodata		     &  $  < 12.77		      $    &14(S)    &  $  < 12.18		$      &  \nodata				   &  \nodata			   & $  18.62 \pm  0.06  $   &          1(CIB)		\\
H\,1821+643		&  $  15.24 \pm 0.02   $	   &  $  24.6 \pm  0.4$ 		 &  $  14.34 :  	$    &  $ 14.34\pm\,^{0.19}_{0.09}\,^a$    &20(S)    &  $  15.75\pm\,^{0.30}_{0.15}$   &  $  18.6\pm\,^{2.9}_{3.1}	   $	   &  $  13.85 \pm 0.05    $	   & $  20.57 \pm  0.06  $   &   3+4+5   			     \\
H\,1821+643		&  $  14.29 \pm 0.08   $	   &  $  6.6 \pm  0.7$  		 &  \nodata		     &  $  < 12.63		      $    &20(S)    &  $  14.14 :		$      &  \nodata				   &  \nodata			   & $  18.95 \pm  0.06  $   &       2(OA)		  \\
H\,1821+643		&  $  14.12 \pm 0.10   $	   &  $  6.3 \pm  1.1$  		 &  \nodata		     &  $  < 12.63		      $    &20(S)    &  $  14.05 :		$      &  \nodata				   &  \nodata			   & $  18.54 \pm  0.06  $   &   	    1(OA)		  \\
HD\,18100		&  $ 14.55 \pm 0.02		 $ &  $  12.2 \pm 0.6		   $	 &  $ 13.88 \pm 0.02	$    &  $  13.95 \pm 0.04	      $    &45(F)    &  $  15.26 \pm 0.04	   $   &  \nodata				   &  $ 13.34 \pm 0.04     $	   & $  20.14 \pm  0.07  $   & (Ly$\alpha$)	 	       \\
\enddata
\end{deluxetable}

\begin{deluxetable}{lcccccccc}
\tabcolsep=2pt
\tablecolumns{9}
\tablewidth{0pt} 
\tabletypesize{\tiny}
\tablecaption{Cooling Rates, Electron Densities, and Depletions \label{t3}} 
\tablehead{\colhead{Target}  &   \colhead{$\log (l_c)_{\rm H}$} &   \colhead{$\log (L_c)_{\rm S}$} &   \colhead{$\log(n_e)_{\rm H}$} &   \colhead{$\log(n_e)_{\rm S}$} & \colhead{[Fe\,{\sc ii}/H\,{\sc i}]} & \colhead{[S\,{\sc ii}/H\,{\sc i}]} & \colhead{[P\,{\sc ii}/H\,{\sc i}]}&\colhead{Components} \\ 
\colhead{(1)}& \colhead{(2)} & \colhead{(3)}& \colhead{(4)}& \colhead{(5)}& \colhead{(6)}& \colhead{(7)}& \colhead{(8)} & \colhead{(9)}}
\startdata
\cutinhead{\sc Low Velocity Clouds}
3C\,273.0		 & $ -25.85 \pm\,^{0.10}_{0.10}  $	 & $ -25.75 \pm\,^{0.04}_{0.04}  $	  & $  -1.32 \pm\,^{0.10}_{0.10}  $    & $  -1.25 \pm\,^{0.04}_{0.04}  $	     & $  -0.78 \pm\,^{0.10}_{0.10}  $      & $  -0.10 \pm\,^{0.10}_{0.11}  $	  & $  -0.25 \pm\,^{0.12}_{0.13}  $   &   1+2+3 	     \\
3C\,273.0		 & $ -25.28 \pm\,^{0.10}_{0.11}  $	 & $ -25.49 \pm\,^{0.05}_{0.05}  $	  & $  -0.75 \pm\,^{0.10}_{0.11}  $    & $  -0.95 \pm\,^{0.05}_{0.05}  $	     & $  -0.46 \pm\,^{0.10}_{0.10}  $      & $   0.21 \pm\,^{0.11}_{0.11}  $	  & $	0.11 \pm\,^{0.14}_{0.15}  $   &   4		   \\
3C\,351.0		 & $ -25.53 \pm\,^{0.66}_{0.19}  $	 & $ -25.68 \pm\,^{0.84}_{0.33}  $	  & $  -1.00 \pm\,^{0.66}_{0.19}  $    & $  -1.15 \pm\,^{0.84}_{0.33}  $	     &  \nodata 			    & $   0.15 \pm\,^{0.76}_{0.25}  $	  &  \nodata			      &   2+3	      \\
H\,1821+643	  	 & $ -25.77 \pm\,^{0.20}_{0.11}  $	 & $ -25.75 \pm\,^{0.33}_{0.19}  $	  & $  -1.24 \pm\,^{0.20}_{0.11}  $    & $  -1.22 \pm\,^{0.33}_{0.19}  $	     & $  -0.83 \pm\,^{0.06}_{0.06}  $      & $  -0.02 \pm\,^{0.30}_{0.17}  $	  & $  -0.28 \pm\,^{0.08}_{0.08}  $   &   3+4+5 		   \\
HD\,18100   	  	 & $ -25.73 \pm\,^{0.08}_{0.08}  $	 & $ -25.65 \pm\,^{0.06}_{0.06}  $	  & $  -1.20 \pm\,^{0.08}_{0.08}  $    & $  -1.12 \pm\,^{0.06}_{0.06}  $	     & $  -1.09 \pm\,^{0.07}_{0.07}  $      & $  -0.08 \pm\,^{0.08}_{0.08}  $	  & $  -0.36 \pm\,^{0.08}_{0.08}  $   & (Ly$\alpha$)		\\
HD\,233622 	  	 & $ -25.64 \pm\,^{0.10}_{0.10}  $	 & $ -25.51 \pm\,^{0.08}_{0.09}  $	  & $  -1.11 \pm\,^{0.10}_{0.10}  $    & $  -0.98 \pm\,^{0.08}_{0.09}  $	     & $  -1.08 \pm\,^{0.08}_{0.08}  $      & $  -0.13 \pm\,^{0.10}_{0.10}  $	  & $  -0.28 \pm\,^{0.09}_{0.09}  $   & 2+3			   \\
HD\,93521  	  	 & $ -25.45 \pm\,^{0.07}_{0.07}  $	 & $ -25.49 \pm\,^{0.03}_{0.03}  $	  & $  -0.92 \pm\,^{0.07}_{0.07}  $    & $  -0.96 \pm\,^{0.03}_{0.03}  $	     &  \nodata 			    & $   0.04 \pm\,^{0.06}_{0.06}  $	  &  \nodata			      &       2+3	       \\
\enddata
\tablecomments{
(1) Target name. 
(2) Logarithmic value of the cooling rate in erg\,s$^{-1}$\, per H\,{\sc i} atom (see Eq.~\ref{ecool1}).
$"<"$: $3\sigma$ upper limit; $">"$: lower limit; ":": uncertain value. 
(3) Logarithmic value of the cooling rate in erg\,s$^{-1}$\, per nucleon (see Eq.~\ref{ecool2}).
$"<"$: $3\sigma$ upper limit; ":": uncertain value. For PG\,1259+593 CIIIC component, the cooling rate 
was corrected for the metal abundance. 
(4) Logarithmic value of the electron density in cm$^{-3}$ using H\,{\sc i} and where the temperature
was set to $T = 6000$ K (see \S~\ref{elecdens} and \S~\ref{elmeas}).  For  HVC Complex C, we did not apply a dust correction, but 
we corrected for the lower metal abundance assuming that the C abundance follows
the abundance of S or Fe (see \S~\ref{hvcgeneral}). 
For VHVC WW84, we also corrected for lower metal abundance (see \S~\ref{hvcgeneral}).
(5) Logarithmic value of the electron density in cm$^{-3}$ using S\,{\sc ii} and where the temperature
was set to $T = 6000$ K (see \S~\ref{elecdens} and \S~\ref{elmeas}).
For Complex C, we did not apply a dust correction. 
(6), (7), and (8) [X\,{\sc ii}/H\,{\sc i}$] \tbond \log(N($X\,{\sc ii}$)/N($H\,{\sc i}$)) - \log({\rm X/H})_\odot$,
where X is either Fe, S, or P. 
(9) The velocity components used to compare the different species from the Gaussian fitting
of the H\,{\sc i} emission profile (see \S~\ref{h1s} and Figs.~\ref{fig2} and \ref{fig2a}).
}
\end{deluxetable}

\begin{deluxetable}{lcl}
\tabcolsep=2pt
\tablecolumns{3}
\tablewidth{0pt} 
\tabletypesize{\scriptsize}
\tablecaption{The C\,{\sc ii} Cooling Rate in Different Environments Obtained from FUV and FIR Observations \label{t5}} 
\tablehead{\colhead{}    &   \colhead{$\log($Cooling Rate)}  & Source}
\startdata
\cutinhead{\sc FUV}
LISM ({\em Copernicus})		&  $-25.46 \pm 0.41 $		& Gry et al. (1992) \\
LISM ({\em FUSE})		&  $-25.63 \pm\,^{0.18}_{0.33}$	& \citet{lehner03} \\
Galactic Halo			&  $-25.70 \pm\,^{0.19}_{0.36}$	& This work \\
IVC (IV Arch)			&  $-25.98 \pm\,^{0.14}_{0.20}$	& This work \\
HVC (Complex C)			&  $-26.99 \pm\,^{0.21}_{0.53}$	& This work \\
DLAs 	 			&  $-26.71 \pm\,^{0.22}_{0.46}$	& Wolfe et al. (2003) \\
\cutinhead{\sc FIR}
Galactic Halo (No CO)		&  $-25.58 \pm 0.10	$	& \citet{bock93} \\
Galactic Halo (CO)		&  $-25.80 \pm 0.11	$	& \citet{matsuhara97} \\
Galactic Halo ({\em COBE})	&  $-25.57 \pm 0.03	$	& \citet{bennett94}  \\
Galactic Halo ({\em ISO})	&  $-25.48 \pm 0.04	$	& \citet{caux97}  \\
\enddata
\tablecomments{For the LISM ({\em Copernicus}) and the IVC (IV Arch), 
the cooling rate is in erg\,s$^{-1}$\, per nucleon.
For the LISM ({\em FUSE}), Galactic Halo, HVC (Complex C), DLAs,
and IR observations, the cooling rate is erg\,s$^{-1}$\, per H atom. 
The given values are the mean cooling rate and the dispersion around
the mean, except for HVC Complex C which consists of only one measurement
toward PG\,1259+593 (hence the error is a $1\sigma$ error on the measurement),
and for the IR observations where it is the best-fit value with its error.
}
\end{deluxetable}

\begin{deluxetable}{lccccc}
\tabcolsep=2pt
\tablecolumns{6}
\tablewidth{0pt} 
\tabletypesize{\scriptsize}
\tablecaption{Estimation of $N($C\,{\sc ii*}$)_{\rm WIM}$ in the WIM  \label{t5a}} 
\tablehead{\colhead{Sightline}    &\colhead{$\log N($C{\sc ii*}$)_{\rm obs}$}   & \colhead{$\Delta \theta$}  &  \colhead{$I[$H\,$\alpha$]}  
&\colhead{$\log N($C{\sc ii*}$)_{\rm WIM}$} & \colhead{$\frac{N({\rm C\,\small II*})_{\rm WIM}}{N({\rm C\,\small II*})_{\rm obs}}$} \\
\colhead{} &\colhead{(dex)} &\colhead{($\degr$)}  &\colhead{(R)} &\colhead{(dex)}&\colhead{}\\
\colhead{(1)}& \colhead{(2)} & \colhead{(3)}& \colhead{(4)}& \colhead{(5)}& \colhead{(6)}}
\startdata
3C\,273.0        &  14.07     &	   0.38     &	 0.83	   &   13.78   &   0.51      \\  
3C\,351.0        &  14.28     &	   0.49     &	 0.95	   &   13.84   &   0.36       \\    
H\,1821+643      &  14.34     &	   0.43     &	 1.85	   &   14.13   &   0.61       \\    
HD\,93521        &  14.03     &	   0.41     &	 0.49	   &   13.16   &   0.13      \\  
HD\,93521 (LVC)  &  13.94     &	   0.00     &	 0.20	   &   13.16   &   0.17       \\  
HD\,93521 (IVC)  &  13.25     &	   0.00     &	 0.15	   &   13.04   &   0.61       \\
HD\,233622       &  14.05     &	   0.18     &	 0.95	   &   13.84   &   0.61       \\
MRC\,2251--178   &  14.17     &	   0.31     &	 0.51	   &   13.57   &   0.25       \\
\enddata
\tablecomments{(2) Total measured C\,{\sc ii*} column density from our observations
(i.e. including all the LVC and IVC components along a sightline). 
(3) $\Delta \theta$ is the distance in degrees to the nearest H$\alpha$  WHAM
survey gridpoint. (4) $I[{\rm H \alpha}]$ intensity in Rayleighs was measured by  WHAM  
\citep{haffner03} over a $1\degr$ diameter field of view. (5,6) Calculated
column density and fraction of C\,{\sc ii*} in the WIM according to Eq.~\ref{eqtwim}. 
See \S~\ref{lvcdiscuss} for more details. 
}
\end{deluxetable}

\clearpage

\begin{figure}[!h]
\begin{center}
\epsscale{0.9}
\plotone{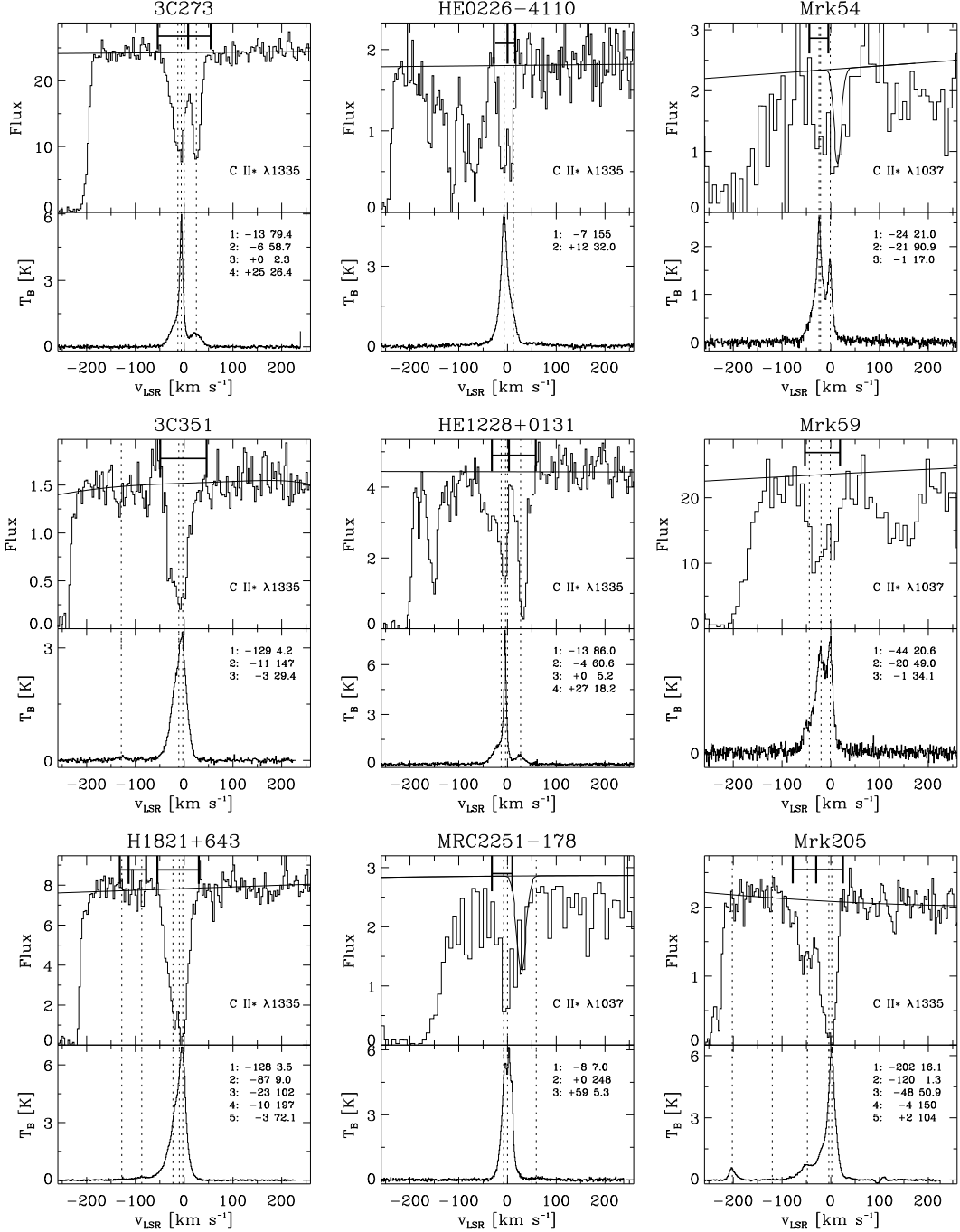}
\caption{\scriptsize Each panel corresponds to one sightline. The bottom sub-panel in each
panel is the H\,{\sc i} brightness temperature as function
of the Local Standard of Rest (LSR) velocity, while 
the top sub-panel shows the flux profiles (in units of $10^{-14}$ erg\,cm$^{-2}$\,s$^{-1}$\,\AA$^{-1}$) 
versus the LSR velocity of the spectrum near C\,{\sc ii*} $\lambda$1037 or $\lambda$1335.  The  vertical dotted lines 
indicate the centroids of the gas cloud components derived from fitting 
Gaussians to the H\,{\sc i} emission profile \citep{wakker01,wakker03}. 
The tick marks in the upper sub-panels indicate the velocity ranges
over which the C\,{\sc ii*} absorption profile was integrated. 
The numbers in each of the bottom sub-panels are from left to right: 
cloud identification number (appearing as well in Tables~\ref{t2} and \ref{t3}); 
the H\,{\sc i} LSR centroid velocity; the H\,{\sc i} column density in 
units of $10^{18}$ cm$^{-2}$. The adopted FUV continuum is shown by the solid line.
The solid absorption line indicates a fit to the H$_2$ absorption line
near C\,{\sc ii*} $\lambda$1037.
}
\label{fig2}
\end{center}
\end{figure}

\begin{figure*}[!h]
\begin{center}
\epsscale{1}
\plotone{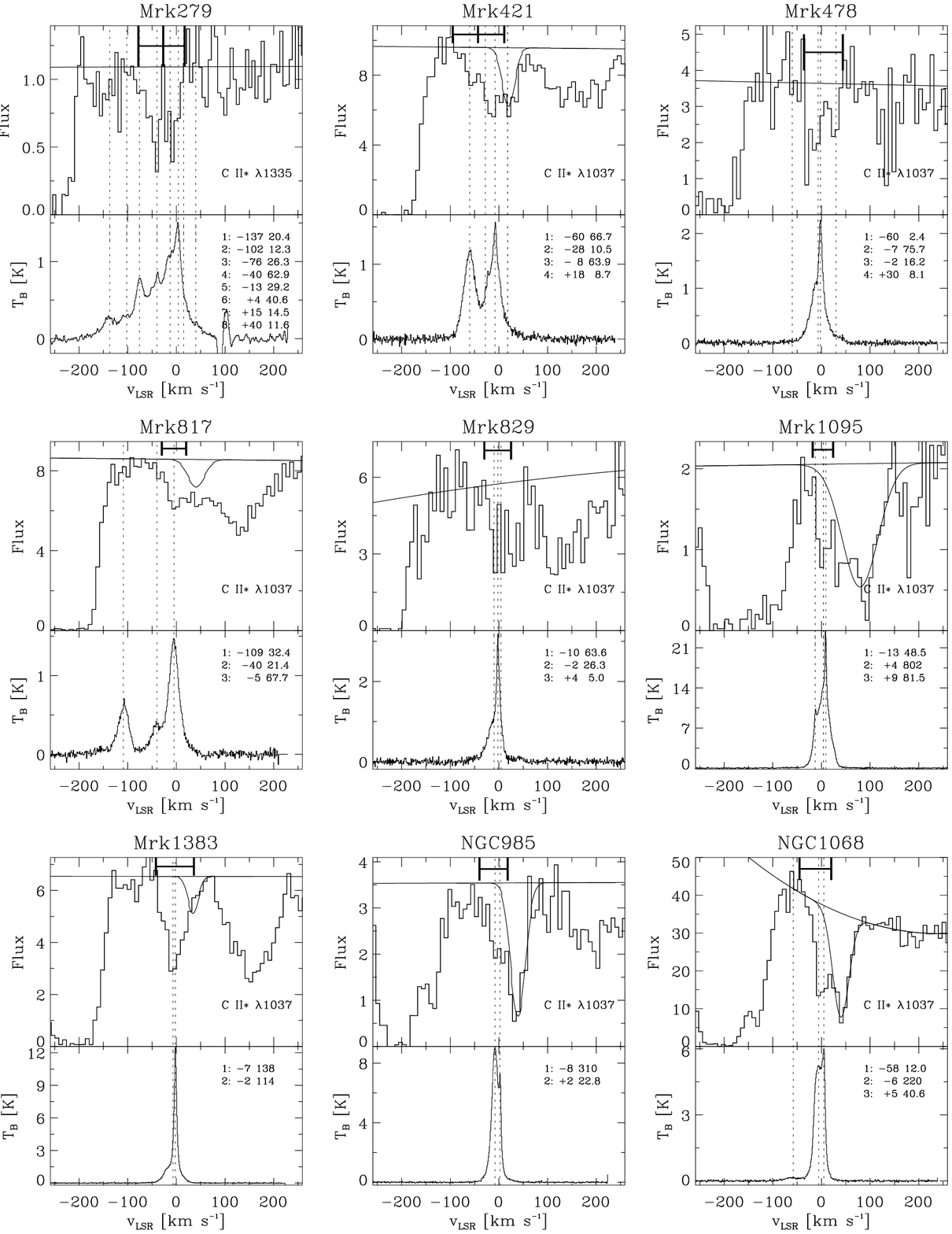}
\figurenum{\ref{fig2}}
\caption{continued.}
\end{center}
\end{figure*}

\begin{figure*}[!h]
\begin{center}
\epsscale{1}
\plotone{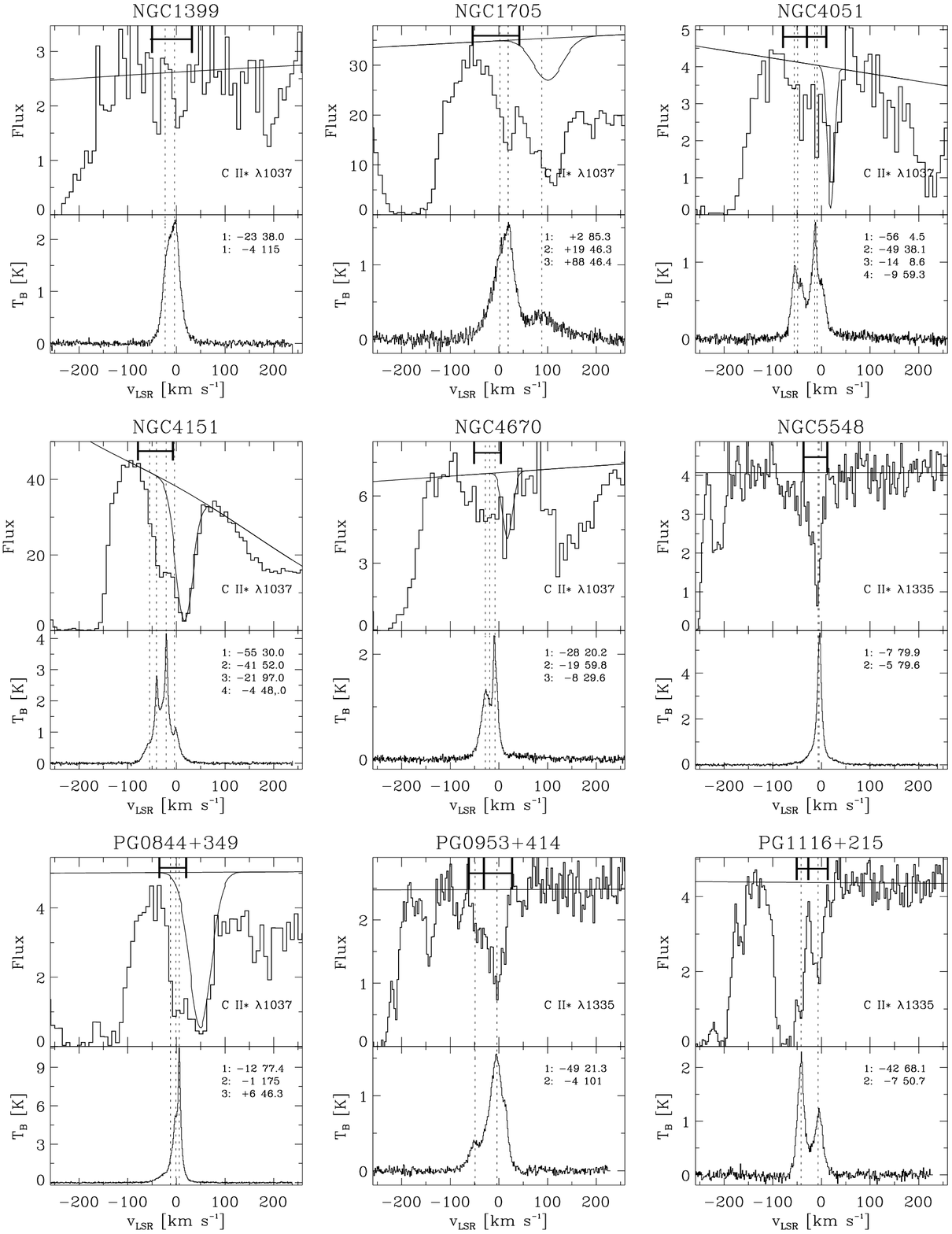}
\figurenum{\ref{fig2}}
\caption{continued.}
\end{center}
\end{figure*}

\begin{figure*}[!th]
\begin{center}
\epsscale{1}
\plotone{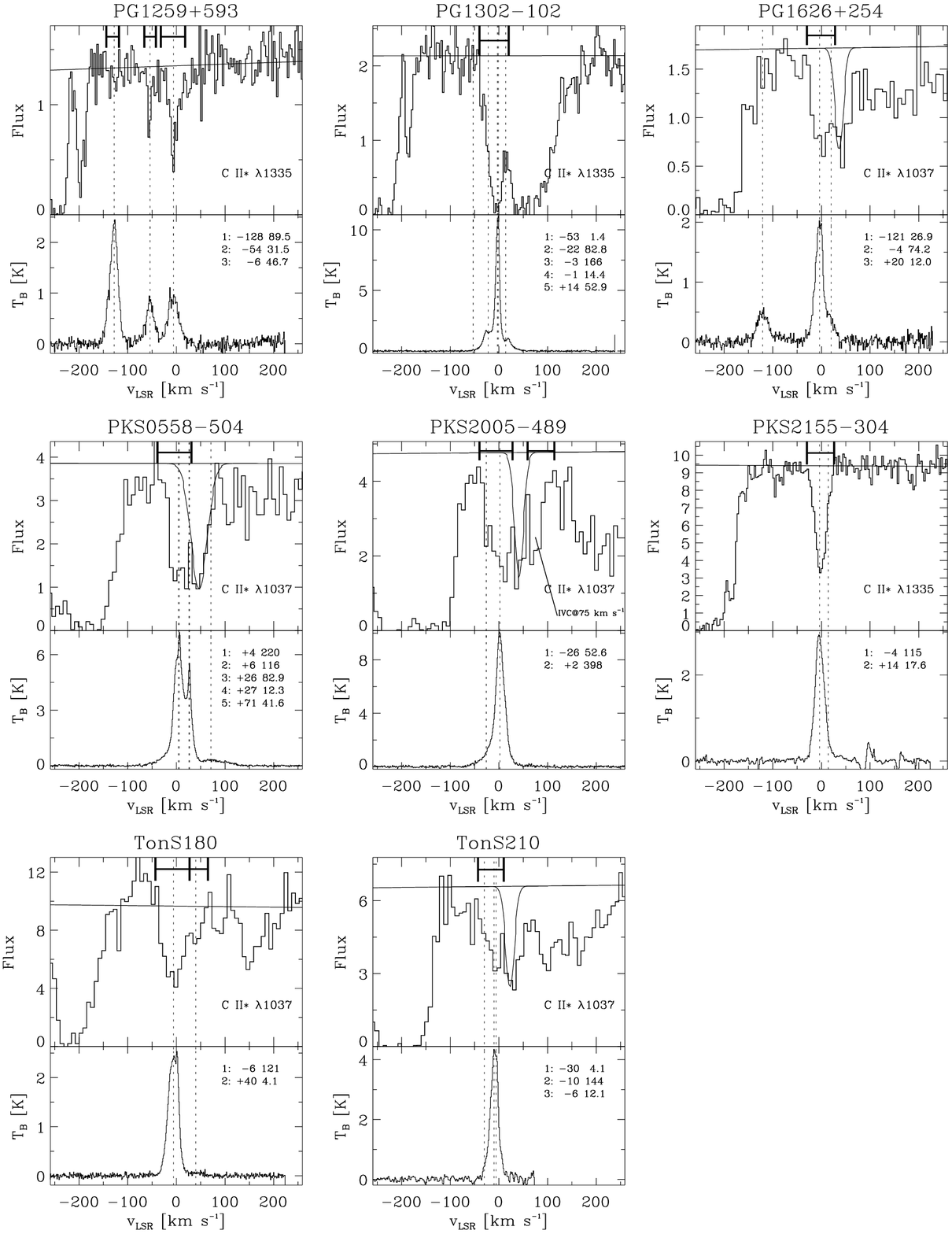}
\figurenum{\ref{fig2}}
\caption{continued.}
\end{center}
\end{figure*}

\begin{figure*}[!th]
\begin{center}
\epsscale{1}
\plotone{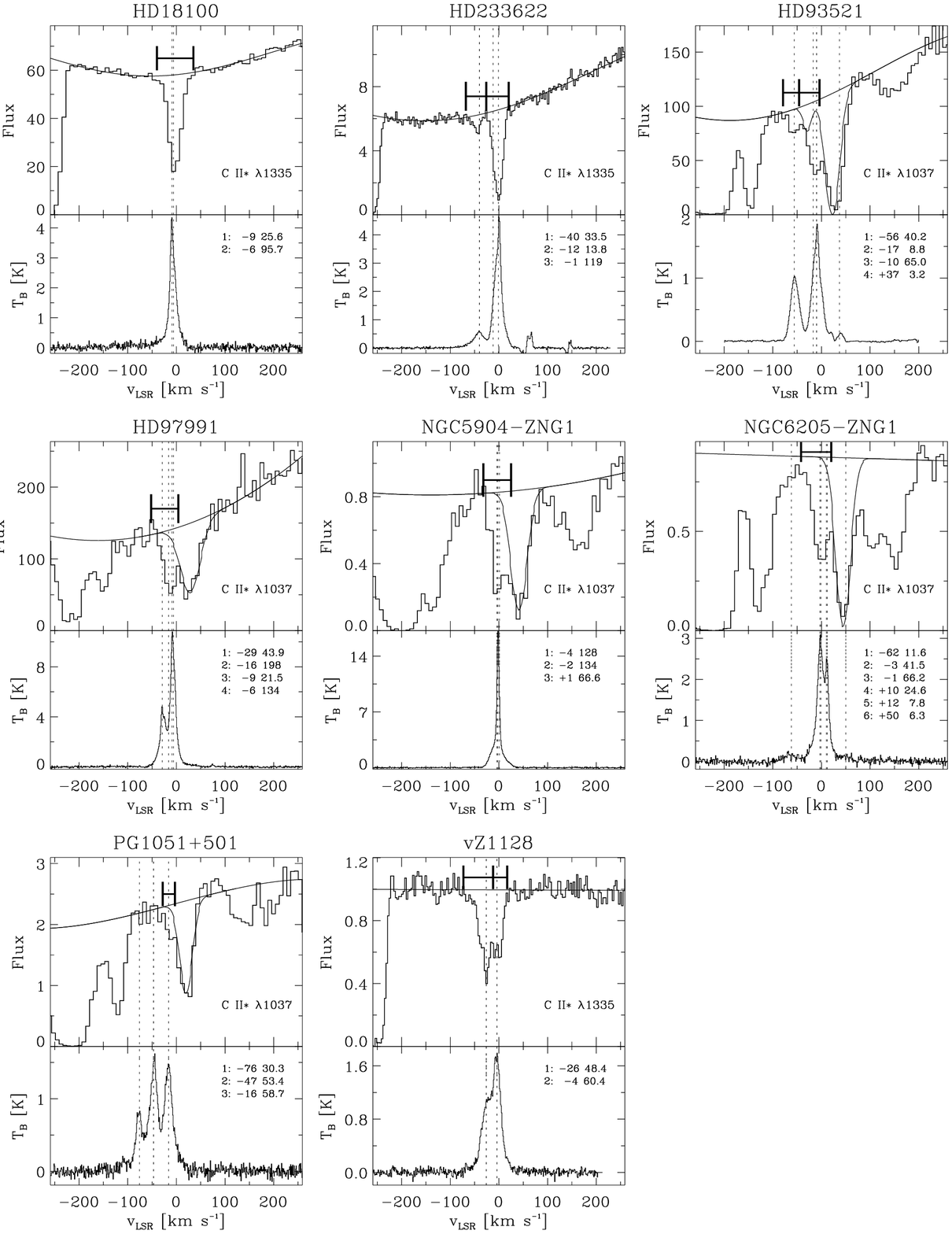}
\caption{Same as for Fig.~\ref{fig2}, but for the stellar sightlines. Here
the fluxes of the FUV spectra are in units of $10^{-12}$ erg\,cm$^{-2}$\,s$^{-1}$\,\AA$^{-1}$. 
For HD\,93521 the fit for H$_2$ includes absorption in the LVC and in the IVC. 
}
\label{fig2a}
\end{center}
\end{figure*}

\begin{figure}[!h]
\epsscale{1}
\plotone{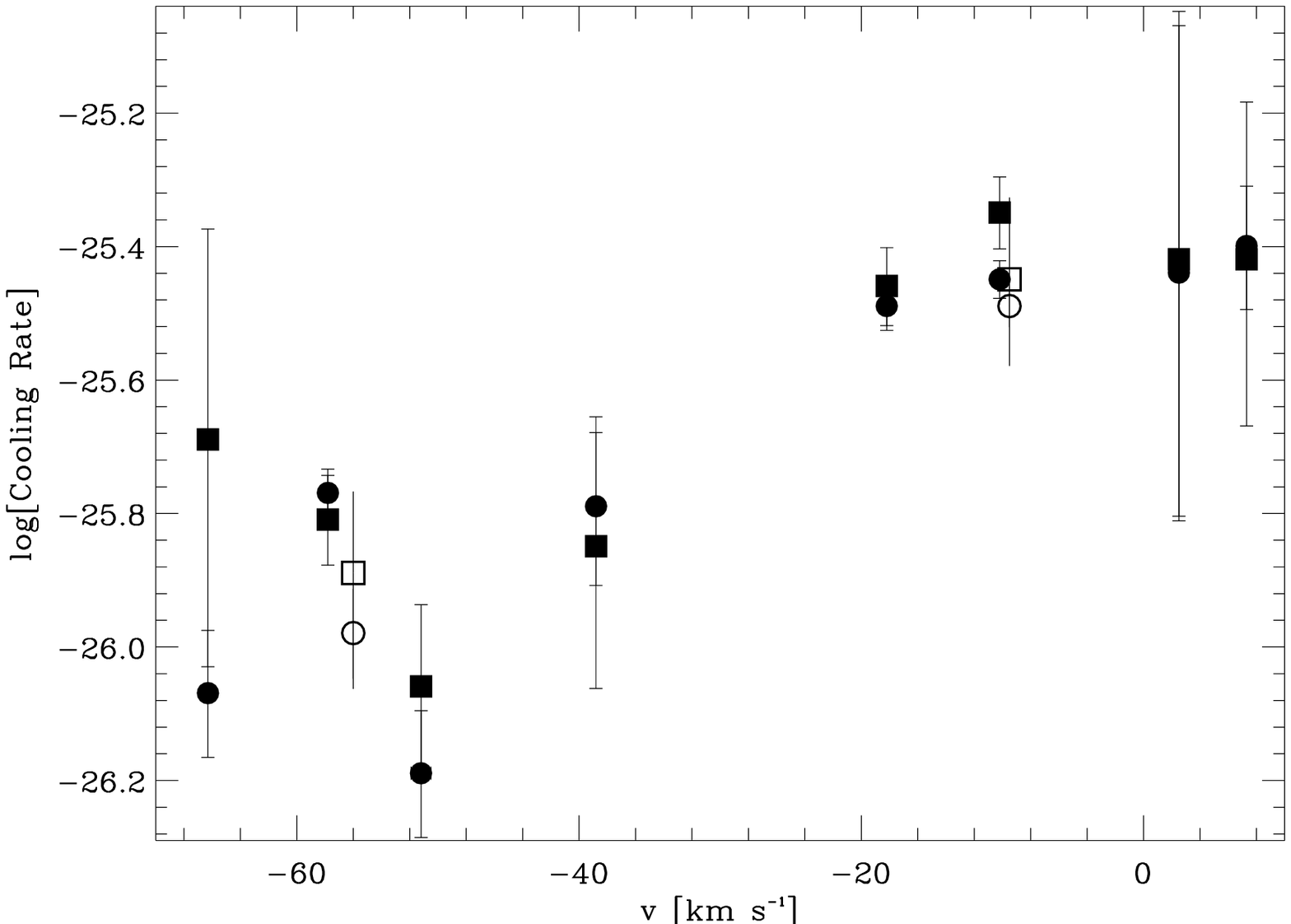}
\caption{Logarithm of the cooling rate in erg\,s$^{-1}$ per H atom (square symbols) 
and erg\,s$^{-1}$ per nucleon (circle symbols) 
against the velocity of each component observed
in the spectrum of HD\,93521. The measurements plotted with filled symbols correspond to the analysis using 3 \km\
resolution GHRS 
observations \citep{spitzer93}. The measurements plotted with open symbols were obtained from 
C\,{\sc ii*} $\lambda$1037 (i.e. from the 20 \km\ resolution {\fuse}\/ data) combined with H\,{\sc i} from the Jodrell Bank telescope. 
\label{hd93}}
\end{figure}

\begin{figure}[!h]
\epsscale{0.8}
\plotone{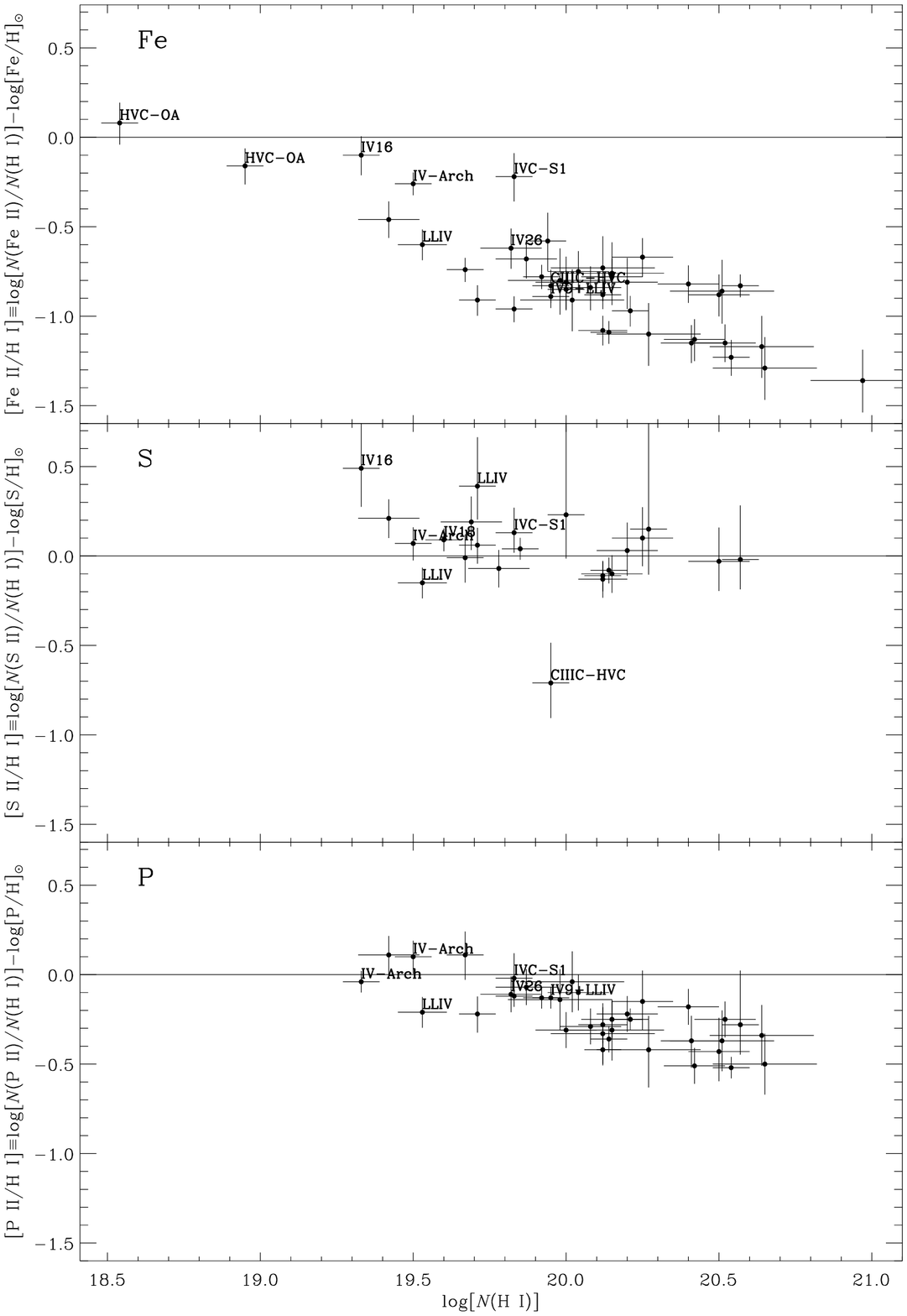}
\caption{Comparison of Fe\,{\sc ii}, S\,{\sc ii}, and P\,{\sc ii} to H\,{\sc i},
normalized to solar abundances.
\label{depl}}
\end{figure}

\begin{figure}[!h]
\epsscale{1}
\plotone{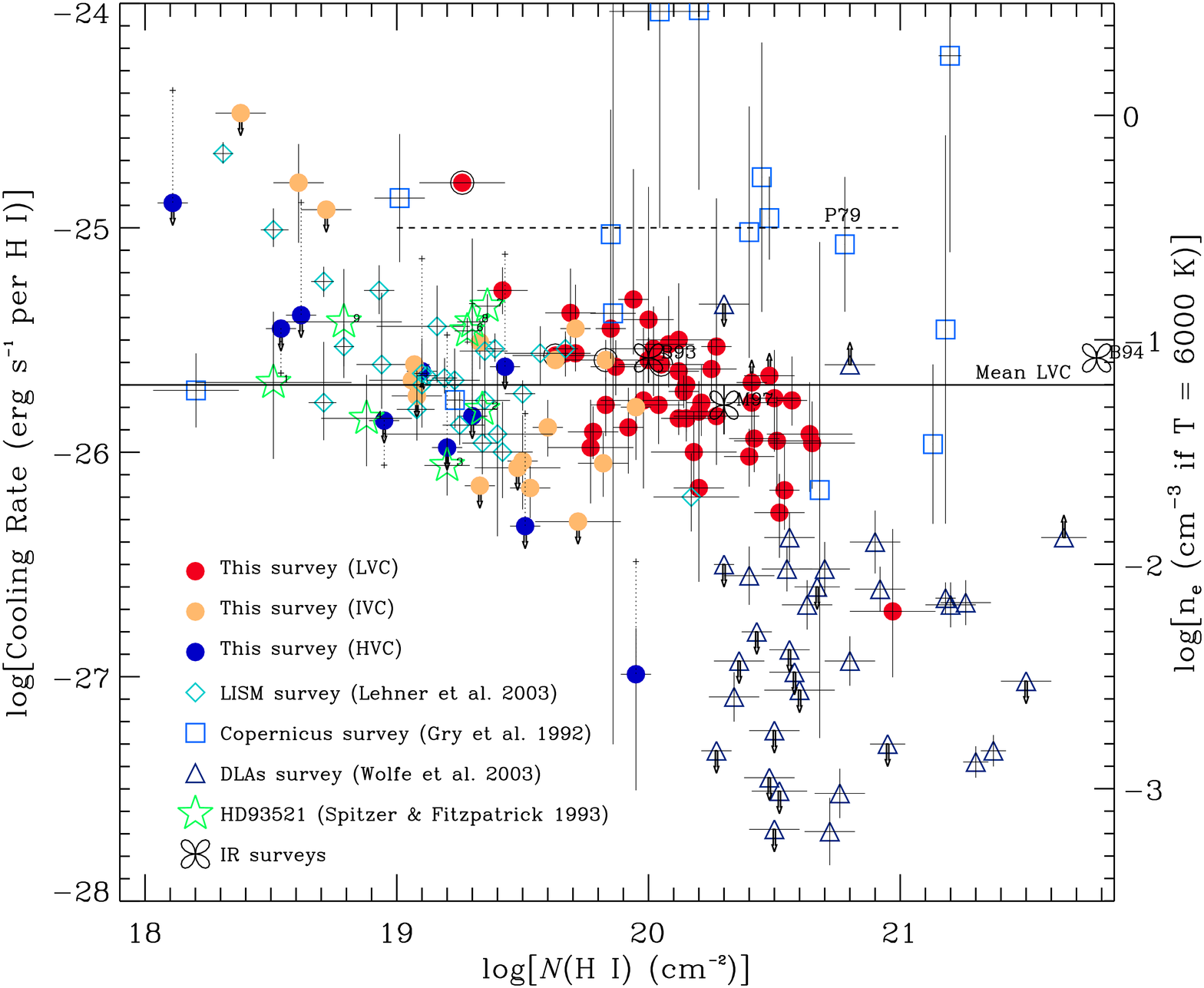}
\caption{\small 
The  logarithm of the cooling rate per H atom (left $y$-axis) is plotted
against the H\,{\sc i} column density from our survey for low-, intermediate-,
and high-velocity clouds. These results are compared to other 
Galactic FUV and high-latitude FIR datasets, and the damped Ly\,$\alpha$ surveys. 
The different results are coded with the symbols given in the legend. 
The cooling rates per H atom of the
values surrounded by an {\em open circle} are uncertain.
For the high-latitude FIR, the results are from Bock et al. (1993, B93, 
for $\log N($H\,{\sc i}$)> 20.00$ dex); Bennet et al. (1994, B94,
for $\log N($H\,{\sc i}$)< 21.78$ dex); Matsuhara et al. (1997, M97, 
for $\log N($H\,{\sc i}$)< 20.30$ dex). The horizontal dashed line
shows the mean cooling rate per nucleon of the study by \citet{pottasch79} 
(their H\,{\sc i} column densities lie between about 19 and 21 dex). 
The mean value of the cooling rates per H atom observed for LVCs in our sample is 
indicated by the solid horizontal line.  
Note that for the \citet{gry92} survey, the cooling rate is actually in erg\,s$^{-1}$\, per nucleon
(and note also that not all their H\,{\sc i} column densities have errors).
The logarithm of the electron density from our survey  
(assuming a temperature of 6000 K and solar abundances) is shown on the right $y$-axis. 
The dotted lines joining some of the data points 
correspond to the correction to $n_e$ for the metallicity and dust (see \S~\ref{elmeas} for more details).
\label{cool}}
\end{figure}

\begin{figure}[!h]
\epsscale{1}
\plotone{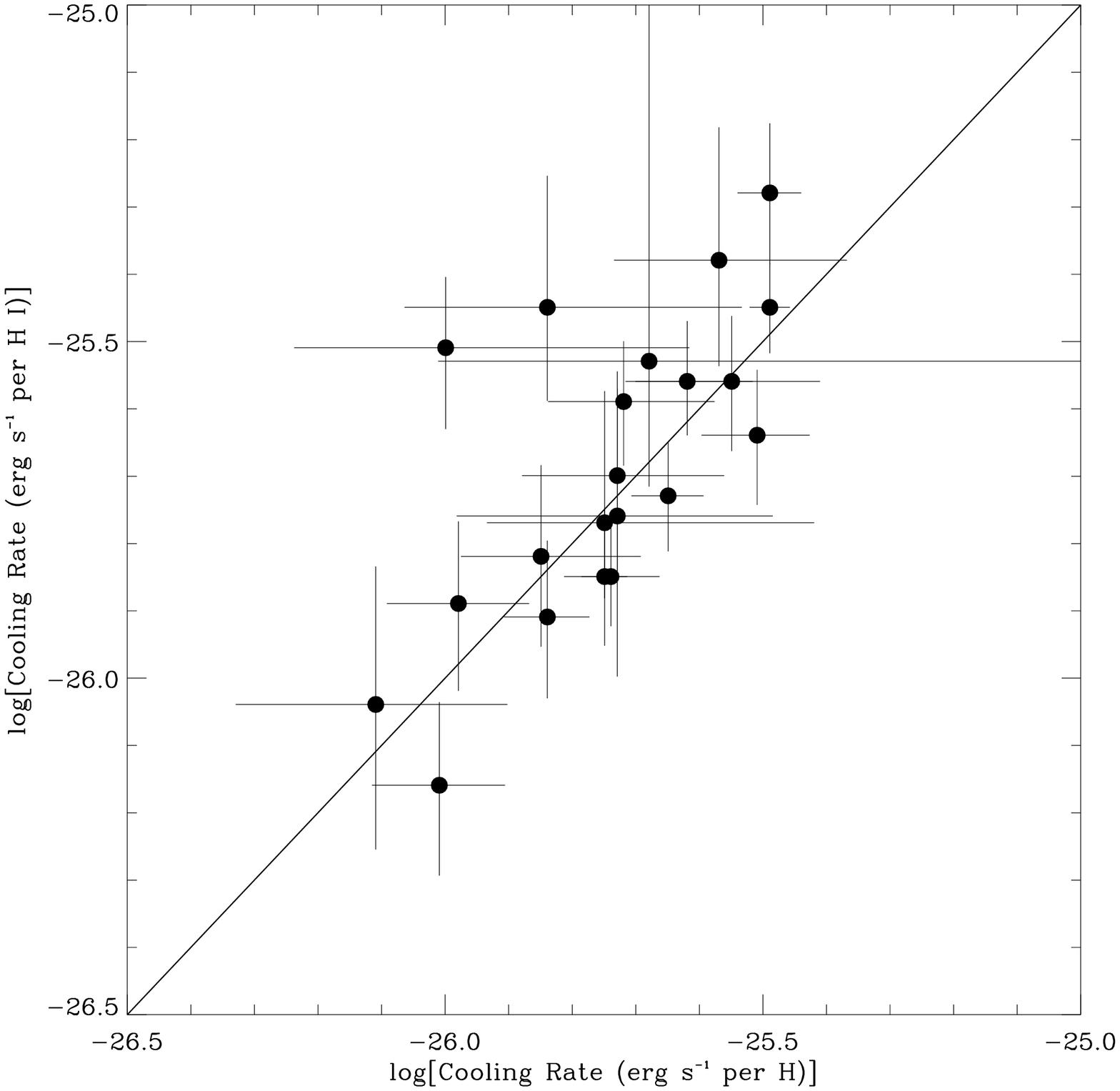}
\caption{Comparison of the cooling rate per H atom  and the cooling rate
per nucleon for LVCs. The cooling rate per nucleon was derived using S\,{\sc ii}. 
The straight line is a 1:1 relationship. See \S~\ref{coolmeas} for more details. 
\label{s2c2}}
\end{figure}

\begin{figure}[!h]
\epsscale{1}
\plotone{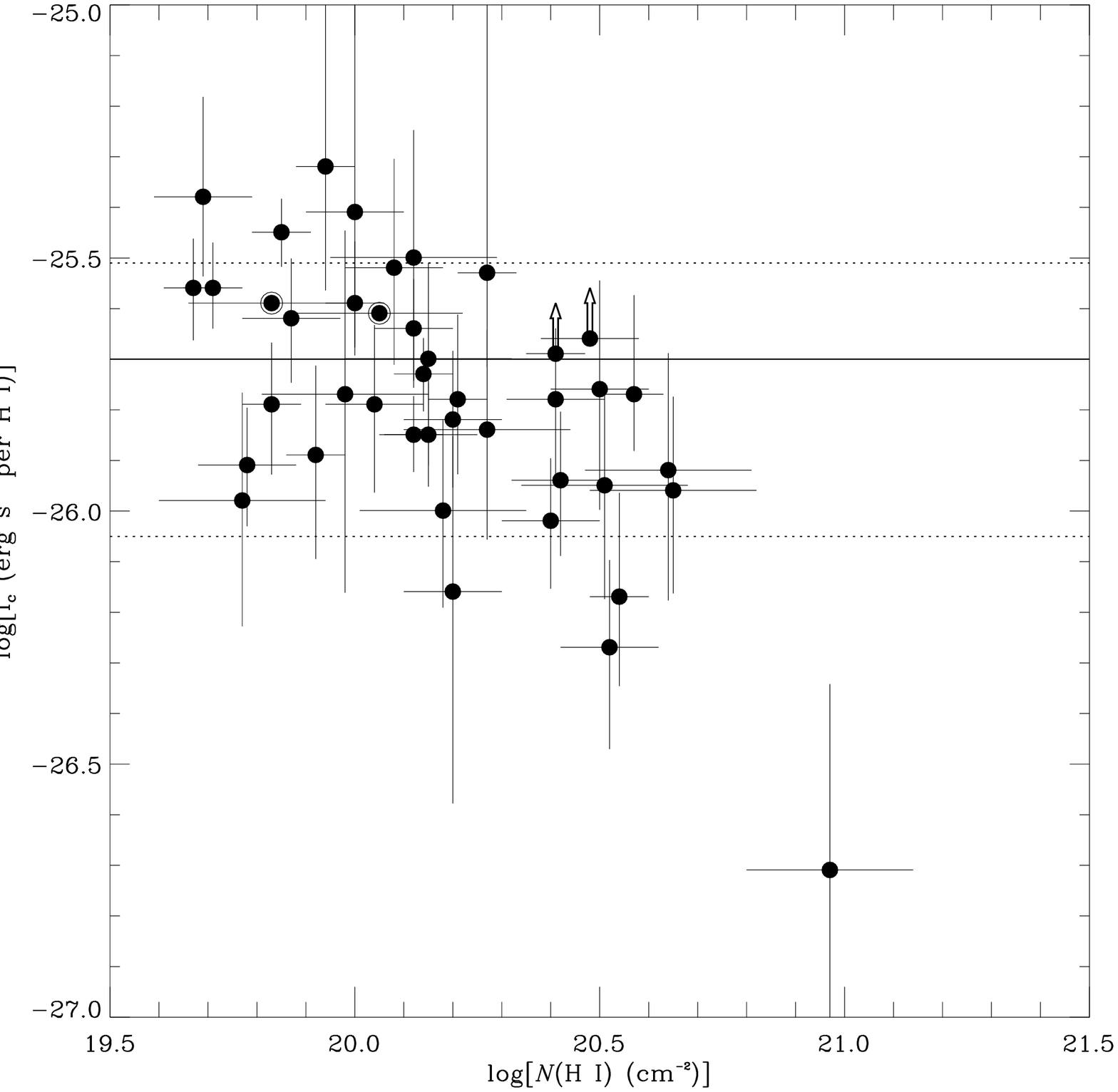}
\caption{The  logarithm of the cooling rate per H atom is plotted
against the H\,{\sc i} column density from our survey for the LVCs. 
The cooling rates per H atom of the
values surrounded by an {\em open circle} are uncertain.
The solid line shows the mean cooling rate of the LVCs and the dotted
lines are the $1\sigma$ dispersion. 
\label{coolbis}}
\end{figure}

\begin{figure}[!h]
\epsscale{1}
\plottwo{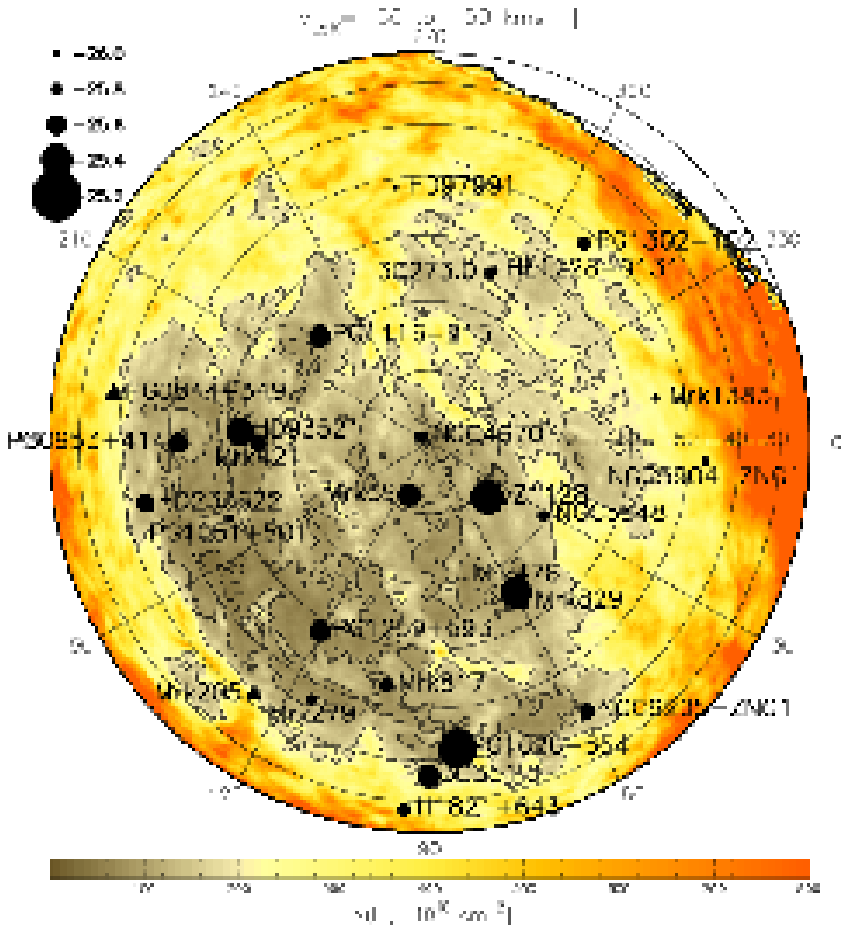}{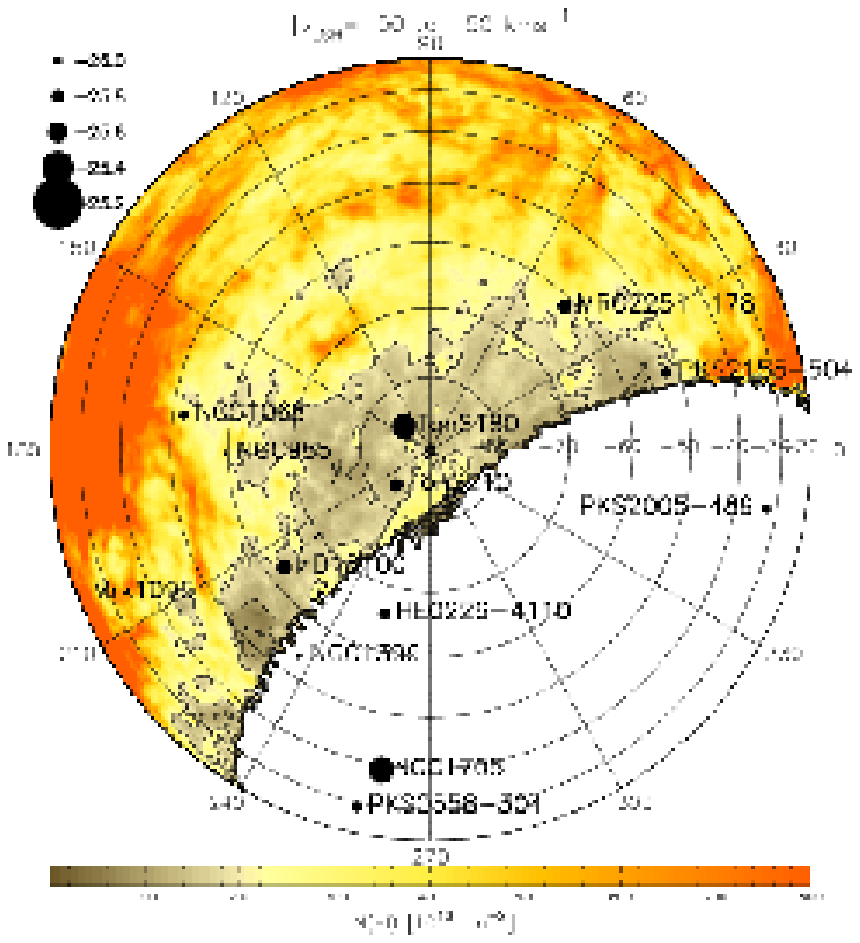}
\caption{H\,{\sc i} contour map of the northern galactic sky (left diagram) and 
southern galactic sky (right diagram) for LSR velocities between $-50$ and
$+50$ \km. The LVC cooling rates per H atom, $l_c$, for each survey object are
displayed with the filled circles. The circle size is proportional to $\log l_c$
according to the legend. A lower limit for PG\,0844+349 is marked with 
a filled triangle. The H\,{\sc i} contour levels are at 5, 20, 50, 100, $200\times 10^{18}$ cm$^{-2}$.
\label{maplvc}}
\end{figure}

\begin{figure}[!h]
\epsscale{1}
\plotone{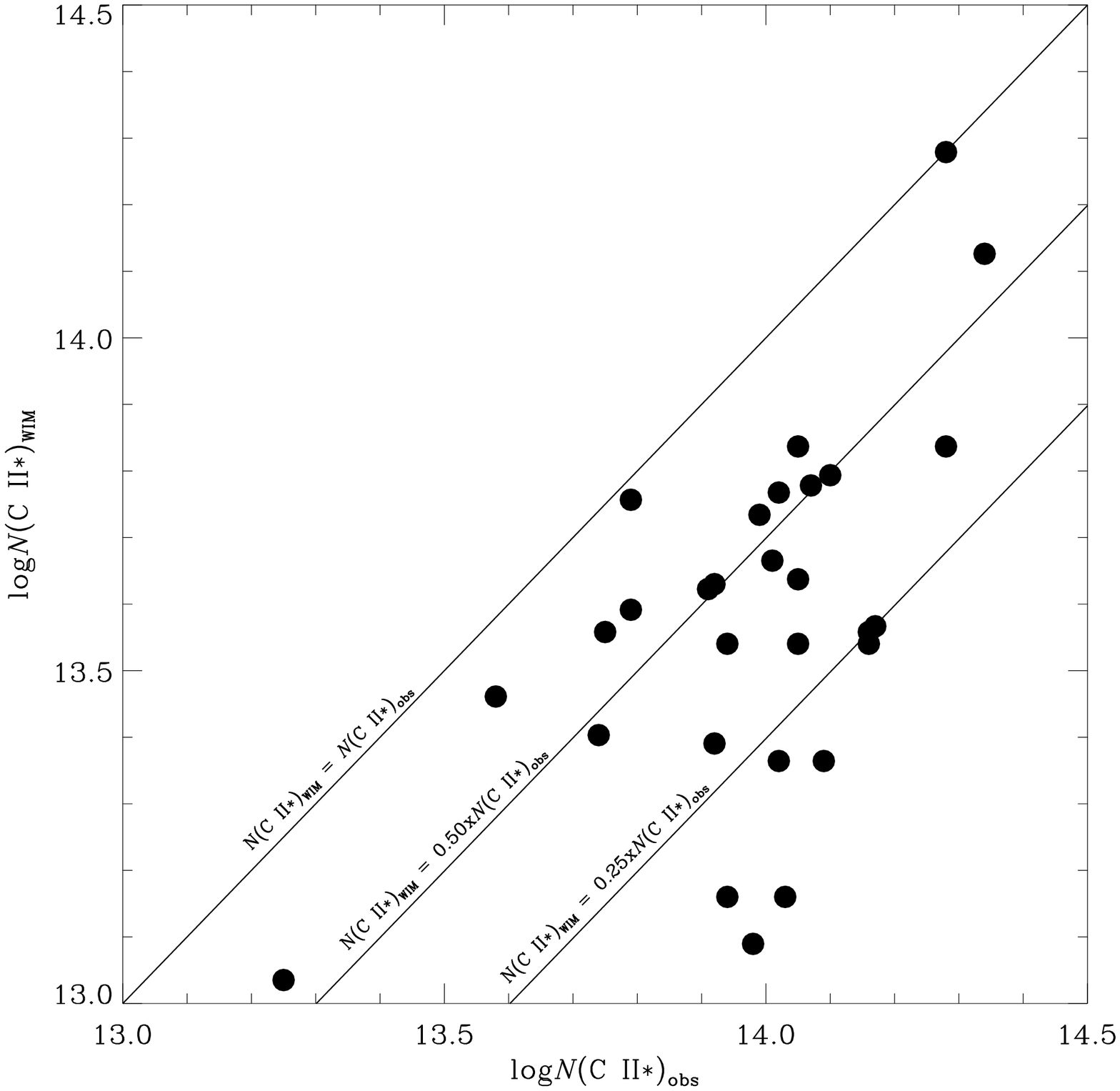}
\caption{The estimated C\,{\sc ii*} column density from the WIM against
the total observed C\,{\sc ii*} column density (see \S~\ref{lvcdiscuss} for
more details).
\label{ciiwim}}
\end{figure}

\begin{figure}[!h]
\epsscale{1}
\plotone{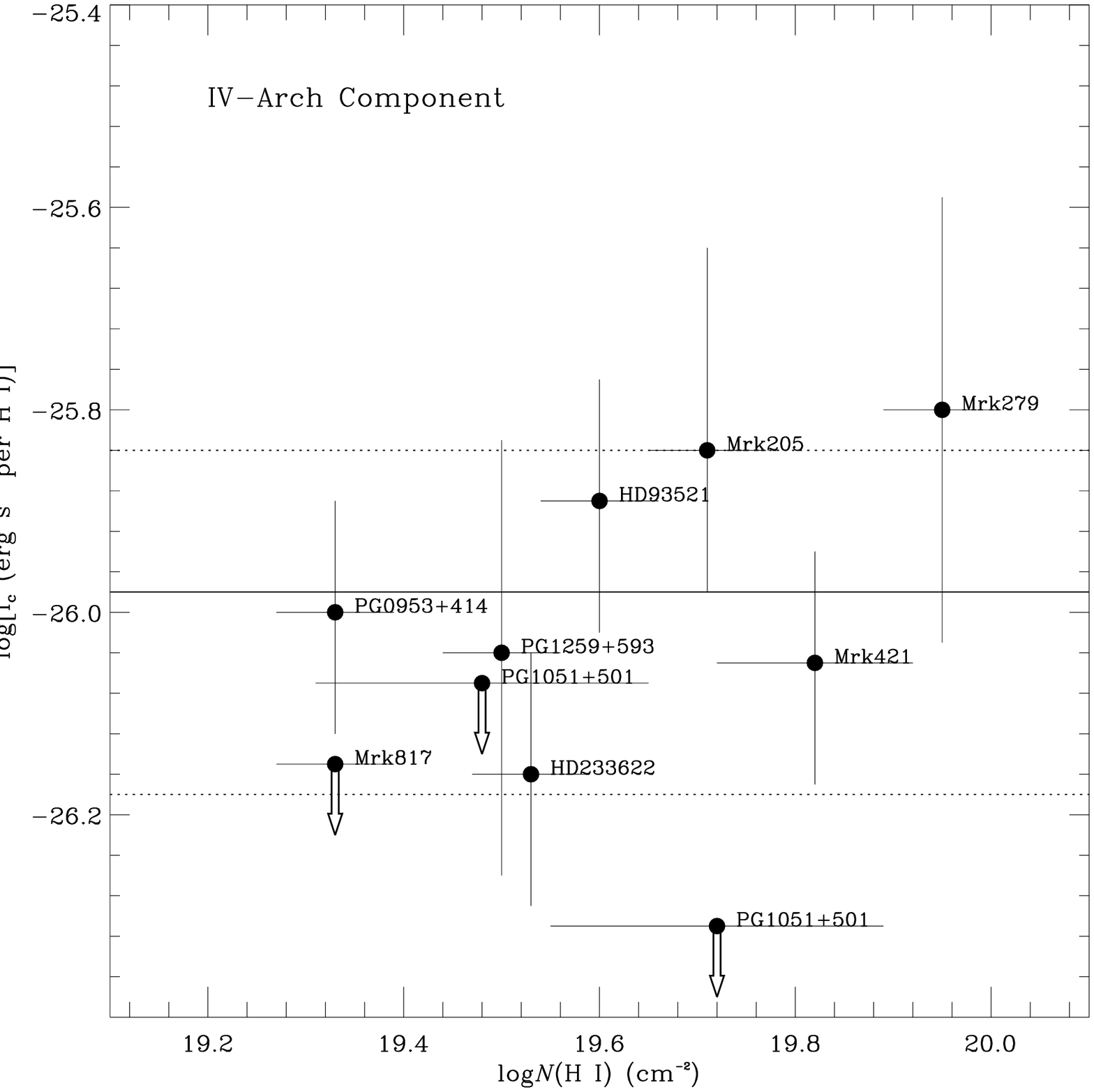}
\caption{The logarithm of the cooling rate against the H\,{\sc i} column density in 
the IV Arch intermediate-velocity cloud. The cooling rate is in erg\,s$^{-1}$ per H atom,
except toward Mrk\,205 and  PG\,0953+414 where it is in erg\,s$^{-1}$ per nucleon,
because a large ionization correction was necessary. For the other sightlines 
the cooling rates per H atom and per nucleon are similar. Note that
the limits toward PG\,1051+501 are uncertain and could be higher because the 
continuum placement is uncertain. The solid line shows the mean cooling rate of the IVCs and the dotted
lines are the $1\sigma$ dispersion. 
\label{ivczoom}}
\end{figure}

\begin{figure}[!h]
\epsscale{1}
\plotone{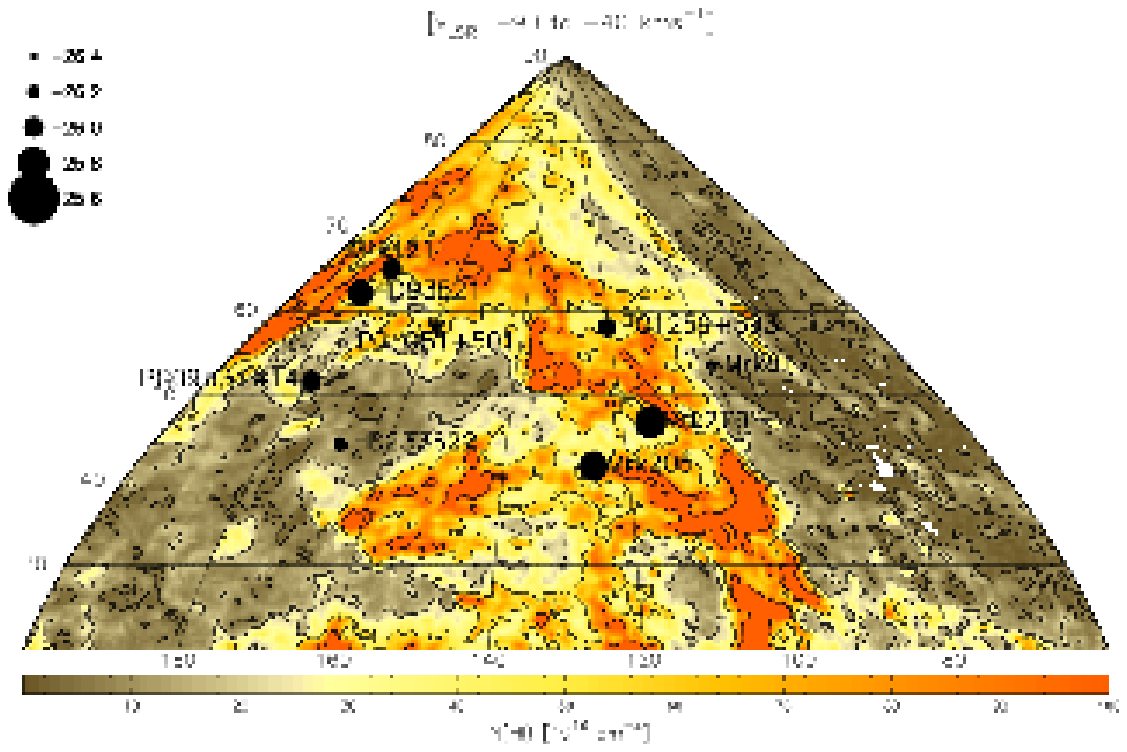}
\caption{H\,{\sc i} contour map of the IV Arch for LSR velocities between $-90$ and
$-40$ \km, where the colors denote the H\,{\sc i} column density. Galactic 
longitude and latitude are displayed on the figure. The logarithms  of the 
cooling rates per H atom derived from the C\,{\sc ii*} UV observations are shown
for each extragalactic or stellar direction as filled circles for actual 
measurements, with circle size proportional to $\log l_c$ (see the legend). Mrk\,59 and NGC\,4151
are not shown because the IVC and LVC absorption are blended. The very 
uncertain measurement for NGC\,4051 is also omitted. A $3\sigma$ upper limit for Mrk\,817 is marked with 
a filled triangle. The H\,{\sc i} contour levels are at 5, 20, 50, 100, $200\times 10^{18}$ cm$^{-2}$.
\label{mapivc}}
\end{figure}

\end{document}